\documentclass[preprint,aps,prd,floatfix]{revtex4-2} 
\pdfoutput=1 
\usepackage{graphicx}
\usepackage[colorlinks=true,
            urlcolor=blue,
            linkcolor=blue,
            citecolor=blue]{hyperref}
\usepackage[T1]{fontenc}
\newcommand{\SH}{Schr{\"o}dinger}
\begin{document} 

\title[Quasinormal modes of a nonsingular spherically symmetric black
hole]{Quasinormal modes of a nonsingular spherically symmetric black
hole effective model with holonomy corrections} 

\author{Douglas M. Gingrich}
\email{gingrich@ualberta.ca}
\altaffiliation[Also at ]{TRIUMF, Vancouver, BC V6T 2A3 Canada} 
\affiliation{Department of Physics, University of Alberta, Edmonton, AB
T6G 2E1 Canada}

\date{\today}

\begin{abstract}
We calculate the quasinormal modes of a nonsingular spherically
symmetric black hole effective model with holonomy corrections.
The model is based on quantum corrections inspired by loop quantum
gravity.
It is covariant and results in a spacetime that is regular
everywhere with a parameter-dependent black bounce. 
Perturbations of these black holes due to massless scalar and
electromagnetic fields have been previously calculated and some
intriguing results were observed.
For some modes, the frequency versus minimum-radius parameter
trajectories were found to spiral and self-intersect in the complex
plane. 
In addition, the spectrum of overtones has real frequencies that
oscillate with increasing overtone number, and may even vanish for
some overtones.
We have calculated the quasinormal modes for all massless spin
perturbations, including spin-1/2, and axial- and polar-gravitational. 
We find that the trajectory-spirals are restricted to scalar
perturbations and observe some interesting overtone behaviour for 
gravitational perturbations.
The amount of isospectrality violation in the gravitational
quasinormal mode spectra is also examined. 
\end{abstract}

\maketitle
\section{Introduction}

In the quest for a theory of quantum gravity, effective black hole
models incorporating quantum corrections provide valuable insights.
The paradigm of loop quantum gravity (LQG) has been useful for
formulating some of these models~\cite{Perez:2017cmj,Zhang:2023yps}.  
The LQG approach has been particularly successful in replacing the
black hole interior singularity with a bounce or transition surface.
Obtaining consistent static exterior solutions that only exhibit
quantum effects in regions of high-curvature is an active area of
research~\cite{ 
Modesto:2008im,
Peltola:2008pa,
Bodendorfer:2019cyv, Bodendorfer:2019nvy, Bodendorfer:2019jay,
Ashtekar:2018lag,Ashtekar:2018cay,Ashtekar:2020ckv,
Gambini:2013ooa,Gambini:2013hna,Gambini:2020nsf,
Kelly:2020uwj,Kelly:2020lec,
Alonso-Bardaji:2021yls,
Alonso-Bardaji:2022ear}.  

We focus on the nonsingular spherically symmetric black hole
effective model with holonomy corrections by Alonso-Bardaji, Brizuela,
and Vera (ABV)~\cite{Alonso-Bardaji:2021yls,Alonso-Bardaji:2022ear}.
The model is based on quantum corrections inspired by loop quantum
gravity.
Anomaly-free holonomy corrections are included through a canonical
transformation and a linear combination of constraints of general
relativity.
The construction is covariant and results in a spacetime that is
regular everywhere with a parameter-dependent black bounce, and two 
asymptotically flat exterior regions. 
The quantum gravity effects introduce a length scale $r_0$.
Curvature scalars are bounded everywhere, and quantum gravity effects
die off with lower curvature. 
Schwarzschild spacetime is recovered for $r_0=0$ and Minkowski
spacetime for vanishing black hole mass. 

One possibility for confronting predicted quantum gravity effects
is to study gravitational mergers.
Gravitational mergers have been observed and can be conceptualised to
occur in three stages: an initial inspiral, the merger, and a ringdown
stage.  
Perturbation theory can be used to gain insight into the ringdown
stage.
A perturbed black hole is a dissipative system and gravitational waves
are emitted as a spectrum of quasinormal modes (QNM) which act as the
spectroscopy of the black hole.
The amplitudes depend on the source of the oscillations, while the
frequencies depend only on the black hole parameters. 
From now on, when we say QNMs, we mean the frequencies of the QNMs, not
the amplitudes. 

Gravitational wave (GW) data from merges is accumulating.
Since the first gravitational waves were
detected~\cite{LIGOScientific:2016aoc,KAGRA:2022twx}, the
LIGO-Virgo-KAGRA Collaborations have recorded 90 GW-burst 
events~\cite{KAGRA:2023pio}.
As a proliferation of possible black hole merger observations is 
anticipated, it is important to study black hole QNMs.

Planned improved sensitivity and efficiency of existing GW 
detectors, as well as the proposed new detectors LISA and the Einstein
Telescope offer a bright future for our ability to deeply probe the
gravitational domain.
Continuous improvements may allow us to constrain new theories of
gravity and possibly one day even shed light on some quantum aspects
of gravity. 

QNMs from scalar perturbations of the ABV quantum corrected metric were
first discussed in \cite{Moreira:2023cxy}, and scalar and vector
perturbations in \cite{Fu:2023drp}. 
In addition, gravitational lensing has been presented in
\cite{Soares:2023uup,Junior:2023xgl}.
Some interesting QNM results have been obtained for scalar
perturbations.
As the distance parameter $r_0$ increases, the QNMs in the
phase-space diagram self-intersect and spiral to a final extremal
value.
This was observed for the first two overtones $n=1,2$ for $s=\ell=0$
only, where $s$ is the spin of the perturbation and $\ell$ is the
azimuthal number. 
No such curves were observed for $\ell = 1$ or $\ell = 2$, for
$n=0,1,2$~\cite{Moreira:2023cxy}.
Spirals were also not observed for electromagnetic
perturbations~\cite{Fu:2023drp}. 
We investigate if these spiral trajectories occur beyond the scalar
$\ell=0$ overtones, and are inherent to the metric or the spin of the
perturbation.

When studying high-overtones of scalar perturbations, the real
part of the QNMs have been observed to oscillate with increasing
overtone number.
This behaviour has been observed for both $\ell = 0$ and $\ell =
1$~\cite{Moreira:2023cxy}.
We address the question if such oscillations also occur for other
spin perturbations.

The QNM spectra from axial- and polar-gravitational perturbations for
asymptotically flat spacetime in general relativity are known to be
identical~\cite{Chandrasekhar,chandrasekharRN,Pani:2013ija,Pani:2013wsa}. 
Such isospectrality is not anticipated for alternative
metrics~\cite{Prasobh:2014zea,
Bhattacharyya:2017tyc,
Bhattacharyya:2018hsj,
Cruz:2020emz,
Chen:2021pxd,
del-Corral:2022kbk}. 
For example, isospectrality is broken for asymptotically AdS black
holes~\cite{Cardoso:2001bb} and has been confirmed in other
works. Asymptotically AdS spacetime is different since the boundary
conditions at spatial infinity are important to recover
isospectrality. This feature has been discussed
in~\cite{Nunez:2003eq,Michalogiorgakis:2006jc,Morgan:2009pn}.
We thus examine the amount of isospectrality violation in the ABV model
for a few overtones and values of $\ell$.

This paper is structured as follows.
In section~\ref{sec:metric}, the effective quantum corrected spacetime
is summarised and stated to have many of the desired features of a
quantum corrected asymptotically flat spacetime.
The perturbation equations, potentials, and coordinates
are introduced in section~\ref{sec:perturbations} for the ABV line
element and for spin $s=0, 1/2, 1,$ and $s=2$ (axial and polar)
perturbations. 
A derivation of gravitational perturbations of general spherically
symmetric static spacetime is outlined in appendix~\ref{app:A}.
The QNM boundary conditions for solving the eigenvalue problem are
given in appendix~\ref{app:B}, along with an asymptotic solution.
The methods employed to calculate the QNMs are introduced in
section~\ref{sec:qnm}.
The results are presented in section~\ref{sec:results} in terms of
QNM phase-space trajectories, higher overtones, and isospectrality
violation.  
For completeness, the first few overtones for all spins are presented
as tables in appendix~\ref{app:C}.
We conclude with a discussion in section~\ref{sec:discussion}.

Throughout, we work in geometric units of $G=c=1$.
Without loss of generality, we take the usual $m=1/2$ $(r_h=1)$ in
numerical calculations.  

%
\section{Effective quantum corrected spacetime\label{sec:metric}}

The ABV model first writes the symmetry reduced Hamiltonian constraint
and diffeomorphism constraint in terms of Asktekar-Barbero variables. 
These are the two densitized triads $\tilde{E}^x$ and
$\tilde{E}^\varphi$, and their conjugate momenta $\tilde{K}_x$ and
$\tilde{K}_\varphi$, where $x$ represents a radial coordinate and
$\varphi$ an azimuthal coordinate.
The holomony corrections are introduced by a polymerization procedure
that replaces the conjugate momenta $\tilde{K}_\varphi$ with a
periodic function $\sin(\lambda K_\phi)/\lambda$. 
The dimensionless parameter $\lambda$ encodes the discretization of
the quantum spacetime and is taken to be positive.
However, to remain anomaly-free in the presence of matter, a canonical
transformation is applied to the densitized triad variables and their
conjugate momenta~\cite{Gambini:2021uzf}:

\begin{equation}
\tilde{E}^x\to E^x\, ,\quad
\tilde{K}_x\to K_x\, ,\quad
\tilde{E}^\varphi\to \frac{E^\varphi}{\cos(\lambda K_\varphi)}\, ,
\quad \textrm{and} \quad 
\tilde{K}_\varphi\to \frac{\sin(\lambda K_\varphi)}{\lambda}\, .
\end{equation}

\noindent
This leaves the diffeomorphism constraint invariant provided
$\cos(\lambda K_\varphi)\ne 0$.

Since the surface $\cos(\lambda K_\varphi)$ may vanish, a 
regularization procedure is applied by defining a linear combination
of the Hamiltonian and diffeomorphism
constraints~\cite{Alonso-Bardaji:2021tvy}.
General relativity is recovered from the new constraint for
$\lambda\to 0$. 

The new constraint algebra gives a structure function which vanishes
at $E^x=0$, like in the Schwarzschild case, and for $\cos(\lambda
K_\varphi)=0$.
From the definition of the constant of motion $m$, one obtains 
$\cos(\lambda K_\varphi) =0$, if, and only if,

\begin{equation}\label{eq:r0}
r_0 \equiv \sqrt{E^x} = 2m\frac{\lambda^2}{1+\lambda^2}\, ,
\end{equation}

\noindent
where $m$ commutes on shell with the modified Hamiltonian.
It is assumed that $m> 0$ and $\lambda \ne 0$, which leads to
$0<r_0<2m$.
The classical theory is recovered in the limit $\lambda\to 0$, which
implies $r_0\to 0$.
The characteristic scale $r_0^2$ arises naturally from the constraint
algebra and defines a minimum area of the model.
The quantity $\cos(\lambda K_\varphi)=0$ now covariantly defines
surfaces on the manifold.

The metric functions are defined in terms of the phase-space
variables in such a way that infinitesimal coordinate transformations
on the spacetime coincide with the gauge variations on the phase
space~\cite{Bojowald:2018xxu}. 
The covariance of the theory means the quantum effects do not depend
on the particular gauge choice, which addresses criticisms raised in 
\cite{Bojowald:2020unm}.
Satisfying these conditions, the line element is

\begin{equation}
ds^2 = -N(r,x)^2 dr^2 + \left( 1 - \frac{r_0}{\sqrt{E^x(t,x)} }
\right)^{-1} \frac{E^\varphi(t,x)^2}{E^x(t,x)} [dx + N^x(t,x)dt]^2 +
E^x(t,x) d\Omega^2\, ,
\end{equation}

\noindent
where $d\Omega^2 = d\theta^2 + \sin^2\theta d\phi^2$ is the standard
Riemannian metric on the unit radius 2-sphere, and $N(t,x)$ and
$N^x(t,x)$ are the lapse and shift functions, respectively,

Different choices of gauge result in distinct charts and their
corresponding line elements for the same metric.
We limit our considerations to the static region of the
quantum-corrected spacetime.
This region is asymptotically flat and describes one exterior region. 
The gauge conditions $K_\varphi = 0$ and $E^x = x^2$ are used to solve
the equations of motion resulting in the ABV line element for the static
region:

\begin{equation}
  ds^2 = -f(r) dt^2 + \frac{1}{g(r)f(r)} dr^2 + r^2d\Omega^2 ,
  \label{eq:metric1}
\end{equation}

\noindent
with

\begin{equation}
f(r) = 1 - \frac{2m}{r} \qquad \textrm{and} \qquad g(r) = 1 -
\frac{r_0}{r}\, ,
\label{eq:metric2}
\end{equation}

\noindent
where the event horizon radius is $r_h = 2m$ and $r_0 < 2m$.
Schwarzschild is restored in the limit $\lambda\to 0$. 

Different geometric interpretations of the mass are possible.
The addition of any function of $r_0$ to $m$ is also a constant of the 
motion.
The constant of motion $m$ is neither the Komar, Hawking
(Misner-Sharp), nor the ADM mass; $m$ is the Komar mass at spatial
infinity and the Hawking mass at the horizon.  
We will express our results in terms of $m$ in correspondence with the 
Schwarzschild expression. 

In addition, the Kretschmann scalar is always positive and finite for
$r>r_0$.
The bounce radius $r_0$ is hidden by the event horizon but
quantum-gravity effects -- parameterized by $r_0$ -- are present 
outside the horizon, and decay as one moves to low-curvature regions.
Some of these quantum-gravity effects are addressed in this paper.

One justifiable criticism of the ABV spacetime is the loss of contact
with a quantum gravity origin~\cite{Borges:2023fub}.
If $\lambda$ is viewed to be a constant over the phase space, the value
of $r_0$ is not fixed and increases as $m$ increases, as opposed to
being fixed at some LQG minimum area.

%
\section{Linear perturbations\label{sec:perturbations}}

In the absence of well defined field equations derived from LQG, the
(test) field equations for massless scalar, electromagnetic, and Dirac
fields are assumed to be the classical field equations in the curved
spacetime whose metric is give by equations (\ref{eq:metric1}) and
(\ref{eq:metric2}). 
We follow the usual procedure when dealing with test fields and
ignore their influence on the background spacetime. This is justified
in first-order perturbation theory since the canonical energy-momentum
tensor is quadratic in the fields and the perturbation field would
contribute only at second-order and higher.
The conservation of these fields may not be satisfied if the matter
sector is not minimally coupled to the metric~\cite{Chen:2019iuo}.  

There is no well defined field equation for quantum spacetimes.
The ABV metric will not be a vacuum solution to the Einstein
equation.
However, we can consider it as a solution to the Einstein equation
with an effective anisotropic matter perfect fluid simulated by the
quantum corrections.
The gravitational perturbations will affect the symmetries of the
background spacetime and the form of the modified Einstein
equation.
We assume at the perturbation level, the quantum corrections are also 
of the anisotropic perfect fluid form.
The derivation of the perturbation equation in~\cite{Yang:2023gas} is
used and sketched in appendix~\ref{app:A}. 

Perturbations $\Psi_s$ of any spin can be written as a
Schr{\"o}dinger-like wave equation 

\begin{equation}
\frac{\partial^2 \Psi_s}{\partial r_*^2} + \left[ \omega^2 -
V_s(r(r_*)) \right] \Psi_s = 0 , \label{eq:schrodinger}
\end{equation}

\noindent
where $V_s$ is a spin-dependent effective potential, $r_*$ is the
tortoise coordinate, and $\omega$ is a complex frequency.

The effective potentials for the ABV metric for $s = 0, 1, 2$ (axial)
and $s=1/2$ can be obtained from the general expressions in
\cite{Arbey:2021jif,Hossenfelder:2012tc,Moulin:2019ekf}: 

\begin{eqnarray}
V_s(r) = f(r)\left[ \frac{\ell(\ell+1)}{r^2} + \frac{(1-s^2)2m -
             (2s+1)(s-1)r_0/2}{r^3}
+ \frac{(2s+3)(s-1)m r_0}{r^4} \right]\, , \label{eq:potential}
\end{eqnarray}

\begin{eqnarray}
V_{1/2}(r) = f(r) \left[ \frac{(\ell+1)^2}{r^2} \pm
      (\ell+1) \sqrt{\frac{r-r_0}{r-2m}} \left( \frac{1}{r^2} 
      - \frac{3m}{r^3} \right) \right] ,
\end{eqnarray}

\noindent
where $\ell \ge s$ for integer $s$ and $\ell \ge 0$ for $s=1/2$.
The case $s=2$ represents axial-gravitational perturbations: $V_2 =
V_2^A$. 
For polar-gravitational perturbations~\cite{del-Corral:2022kbk},

\begin{equation}
V_2^P(r) = \frac{f(r)}{6r^4} \frac{A(r)}{[r(4Lr+5r_0)+2m(6r-7
    r_0)]^2}\, , 
\end{equation}

\noindent
where

\begin{eqnarray}
A(r) & = & r^3\left( 192L^3r^3+48L^2r^2(4r+5 r_0) + 300Lr
r_0^2+125r_0^3\right)\nonumber\\ 
& & + 12m^2r(6r-7r_0)^2(4Lr+5r_0)\\
& & + 6mr^2(6r-7r_0)(4Lr+5r_0)^2 + 8m^3(6r-7r_0)^3\, ,\nonumber
\end{eqnarray}

\noindent
and $L=\ell(\ell+1)/2-1$.

For the ABV metric, the tortoise coordinated in
(\ref{eq:schrodinger}), for $r_0\ne 2m$, is obtained from

\begin{eqnarray}
\frac{dr_*}{dr} & = & \frac{1}{\sqrt{1 - \frac{r_0}{r}} \left( 1
  -\frac{2m}{r} \right)}\nonumber\\
&  = & \sqrt{ 1 - \frac{r_0}{r}} \left[ 1 +
  \frac{4m^2}{(r-2m)(2m-r_0)} - \frac{r_0^2}{(r-r_0)(2m-r_0)}
  \right]\, . 
\end{eqnarray}

\noindent
The former expression  -- actually, its reciprocal -- will be used as
the Jacobian in changes of variable.
The later expression is a useful form for integration.

Less well defined is the integration to obtain $r_*$ as a
function of $r$, since the constant of integration is arbitrary.
We use the following integral

\begin{eqnarray}
r_*(r) & = & \sqrt{1-\frac{r_0}{r}}\ r + \left( 2m + \frac{r_0}{2}
\right) \ln\left[ \left( 1 + \sqrt{1-\frac{r_0}{r}}\, \right)^2
\frac{r}{4}\right]\nonumber\\
& & - \frac{2m}{\sqrt{1-\frac{r_0}{2m}}} \ln \left[
  \frac{\left( 1 + \sqrt{\left(1 - \frac{r_0}{r} \right)\left(1 -
      \frac{r_0}{2m} \right)}\, \right)^2 2 m r - r_0^2}{\left( 1 +
    \sqrt{1-\frac{r_0}{2m}}\, \right)^2 2m(r-2m)} \right] .
\end{eqnarray}

\noindent
Constants have been included in the expression to reproduce
the common Schwarzschild result for $r_0=0$, and to have no constants
in the $r\to \infty$ and $r\to 2m$ limits, reproducing the results
below.
The complete $r_*(r)$ function is only needed when plotting the
potentials as a function of $r_*$; thus the potentials in $r_*$
include an arbitrary constant. 

The asymptotic tortoise relation and coordinate can also be
determined without knowing the exact form of the tortoise coordinate.
For $r\to\infty$,

\begin{equation}
\frac{dr_*}{dr} \to \left( 1 + \frac{r_0}{2r} \right) \left( 1 +
\frac{2m}{r} \right) \approx 1 +\frac{2m+r_0/2}{r}\, ,
\end{equation}

\begin{equation}
r_* \approx r + \left( 2m + \frac{r_0}{2} \right) \ln r\, .
\end{equation}

\noindent
For $r\to 2m$,

\begin{equation}
  \frac{dr_*}{dr} \to \frac{2m}{\sqrt{1-\frac{r_0}{2m}}} (r-2m)^{-1}\, ,
\end{equation}

\begin{equation}
  r_* \approx \frac{2m}{\sqrt{1-\frac{r_0}{2m}}} \ln(r-2m)\, .
\end{equation}

\noindent
These expressions will be useful when determining asymptotic solutions.

\begin{figure*}[htb]
\includegraphics[width=\linewidth]{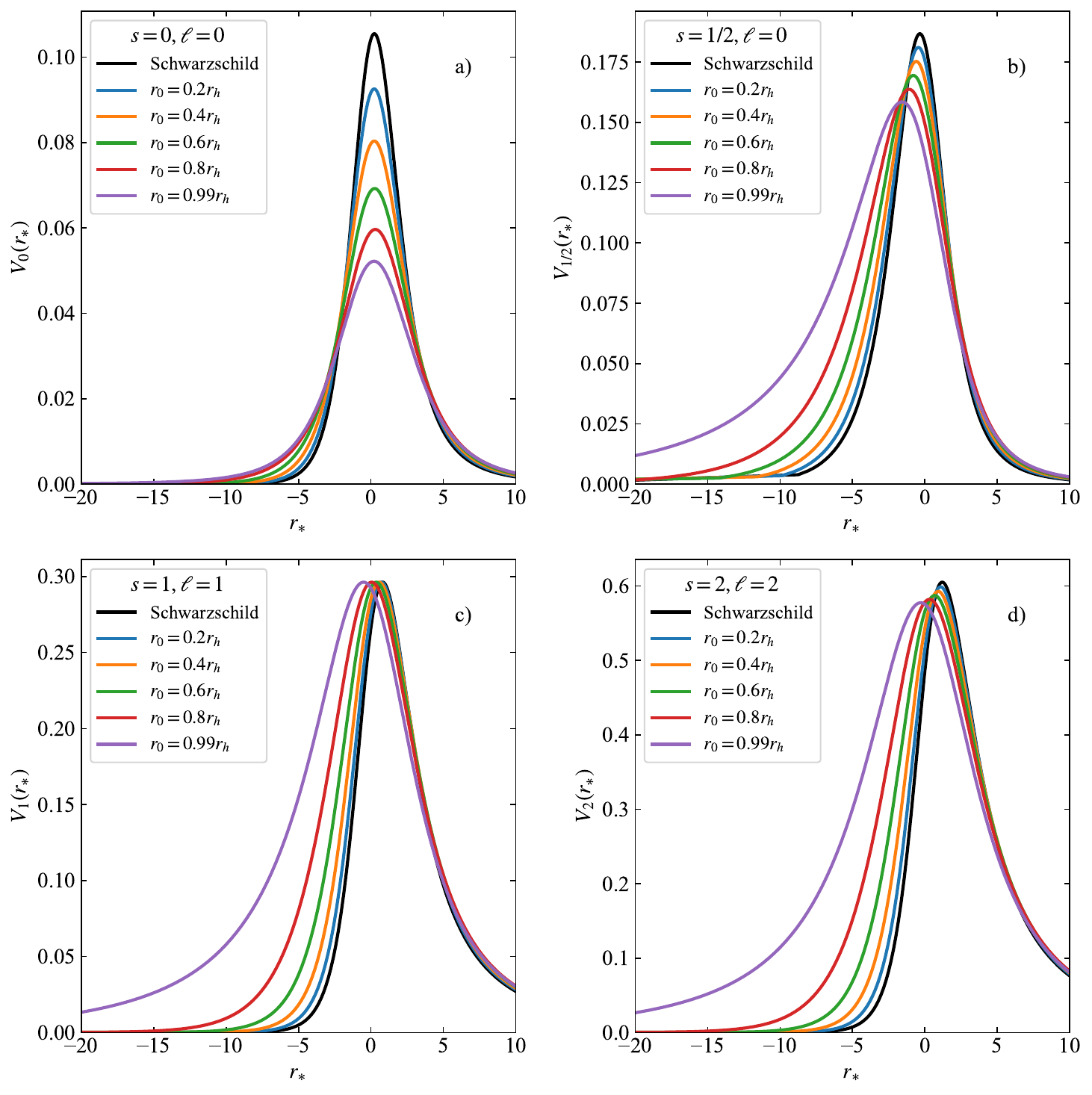}
\caption{\label{fig:V0}Potentials versus $r_*$ for a) spin-0, b)
  spin-1/2, c) spin-1, and d) spin-2 (axial) perturbations on the
  holonomy corrected black hole background metric for different values
  of $r_0$ and lowest values of $\ell$. A mass $m=1/2$ has been used.} 
\end{figure*}

\begin{figure*}[htb]
\includegraphics[width=\linewidth]{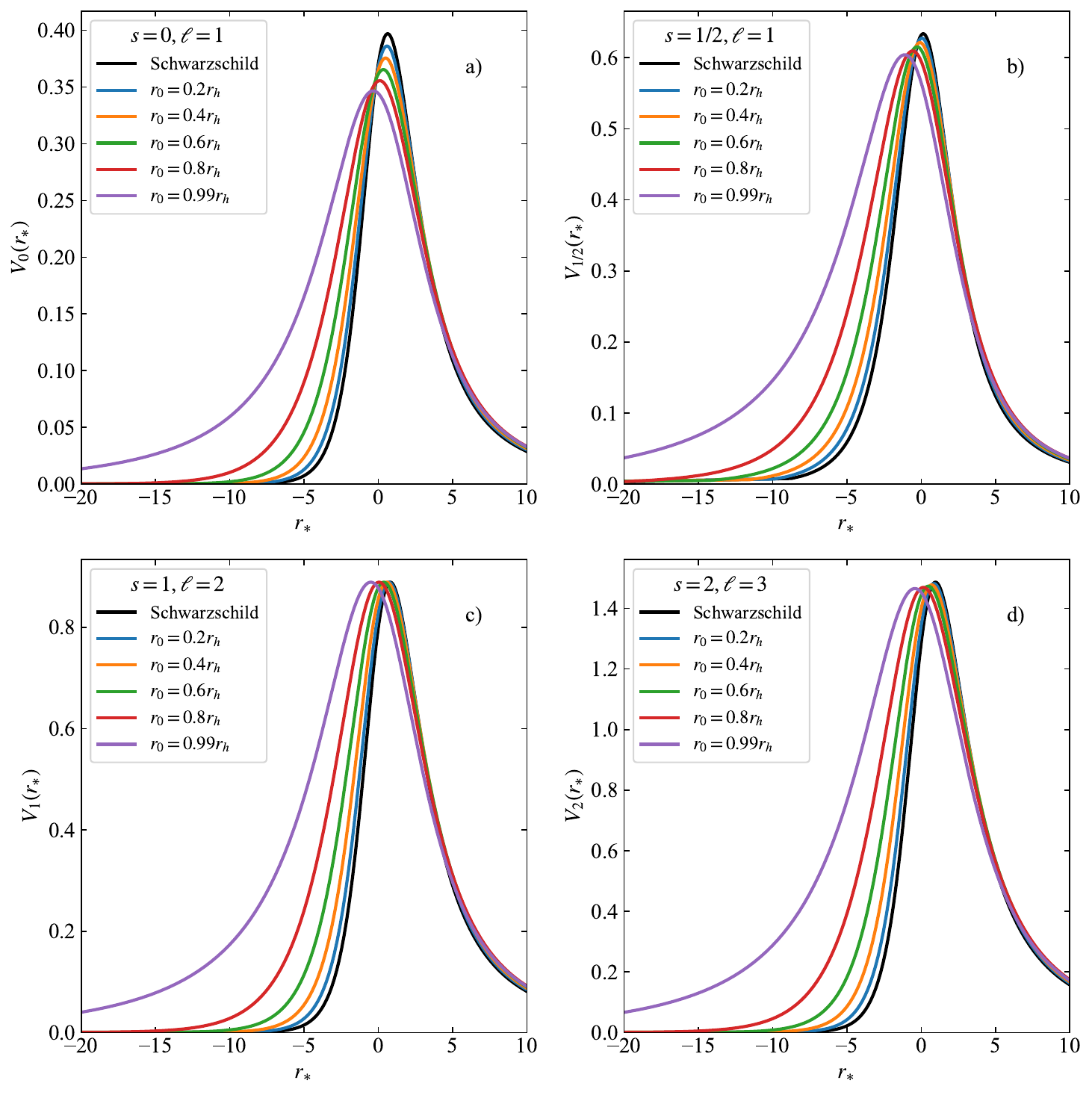}
\caption{\label{fig:V1}Potentials versus $r_*$ for a) spin-0, b)
  spin-1/2, c) spin-1, and d) spin-2 (axial) perturbations on the
  holonomy corrected black hole  background metric for different
  values of $r_0$ and higher values of $\ell$. A mass $m=1/2$ has been
  used.}  
\end{figure*}

The potentials for various values of $r_0$ and the lowest values of
$\ell$ are shown in figure~\ref{fig:V0}; the next higher values of $\ell$
in figure~\ref{fig:V1}.  
Direct comparison with previous results for $s=0$ and $s=1$ are not
possible as those works do not state their expression for $r_*$.
We notice the $s=\ell=0$ potentials appear distinct due to the absence
of the centrifugal term $\ell(\ell+1)/r^2$.
The $s=1$ potentials have the same maximum value since the potentials
are independent of $r_0$, which only appears in the tortoise
coordinate. 
The slow drop in the low-$r_*$ tail of most potentials is due to the
high $r_0$ in the tortoise coordinate, which actually diverges for
$r_0=r_h$. 
For $s=\ell=0$ and high values of $r_0$, the lower-$r_*$ tail in the
potential falls faster than the other potentials due to the
approximate cancellation of terms in the square bracket of
(\ref{eq:potential}). 

\begin{figure*}[htb]
\includegraphics[width=\linewidth]{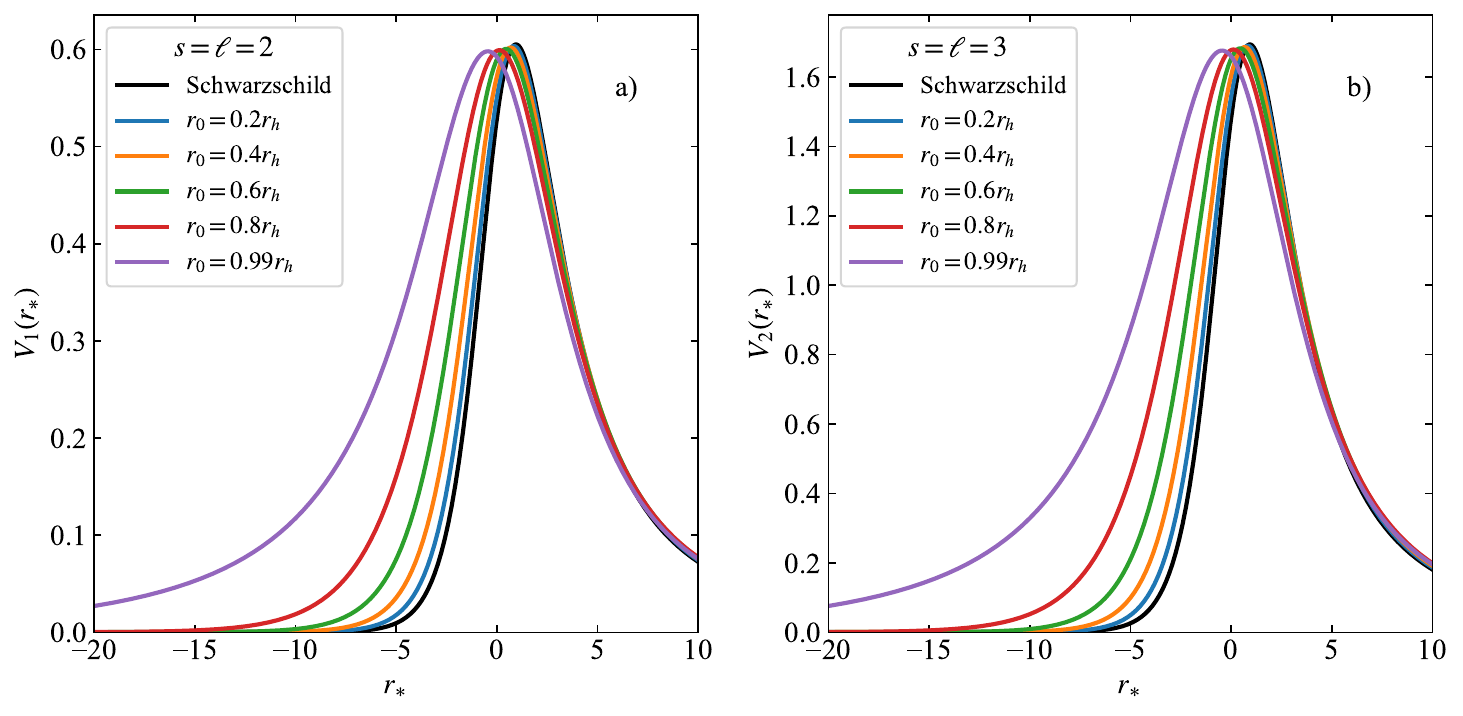}
\caption{\label{fig:VP}Polar-gravitational potentials versus $r_*$ for
  perturbations on the holonomy corrected black hole background metric:
  a) $\ell = 2$ and b) $\ell = 3$. A mass $m=1/2$ has been used.} 
\end{figure*}

Figure~\ref{fig:VP} shows the polar-gravitational potentials.
As expected, the axial- and polar-gravitational potentials differ by
only a small amount.

%
\section{Quasinormal mode frequency calculations\label{sec:qnm}}

We will take advantage of three different methods for calculating
QNMs.
The semi-analytical Wentzel-Kramers-Brillouin (WKB) method which we
use for all spin perturbations, but is limited in accuracy for
high-$n$ and low-$\ell$.  
The pseudo-spectral method (PSM) will be used for low overtones, while
the continued fraction method (CFM) will be employed for calculating
higher overtones.

\subsection{Wentzel-Kramers-Brillouin method}

The WKB method was first applied to the calculation of QNMs in
\cite{1985ApJ...291L..33S}, then developed to 3'rd-order in
\cite{PhysRevD.35.3621}.
Although not particularly accurate for overtones, the method has the
advantage of being semi-analytic and only requiring derivatives of the
potential at the maximum.

The WKB method uses the following generic formula to calculation the
QNM frequencies to order $p$:

\begin{equation}
\omega^2 = V_0 - i \sqrt{-2 V_0^{\prime\prime}} \left[ n + \frac{1}{2}
  + \sum_{i=2}^{p}\Lambda_i \right], \ \ n=0,1,\ldots\, ,
\end{equation}

\noindent
where $V_0$ and $V_0^{\prime\prime}$ are $V(r_*)$ and its second
derivative calculated at the maximum of the potential, and $n$ is
the overtone number. 
Each successive order contributes a term $\Lambda_i$ which is a
function of higher derivatives of the potential evaluated at the
maximum of the potential. 
Each successive term contributes to either the real or imaginary part
of $\omega^2$. 
Higher order does not necessarily give better results.

In order to improve the WKB method, we employ the Pad\'{e}
approximation~\cite{Matyjasek:2017psv}.
The 13'th-order WKB code with Pad\'{e} approximation of
\cite{Konoplya:2019hlu} is used.
We take the lowest relative uncertainty estimate to determine the
optimal order of the calculation. 

\subsection{Pseudo-spectral method}

In this section, we describe how we used the PSM to obtain QNMs.
Spectral methods for solving differential equations are, in general,
powerful and efficient provided the function is smooth, such as our case.
The method approximates the solution we are trying to find, rather
than the equation to be solved.

The \SH-like equation~(\ref{eq:schrodinger}) in $r_*$ is useful for
applying the QNM boundary conditions, but its unbounded nature is not
natural for numerical computations. 
We make a change of variable $u = 2m/r$ to restrict the region outside
the black hole $2m \le r < \infty$ to $0 < u \le 1$. 
In addition, following the approach of \cite{Mamani:2022akq}, we
applying the transformation 

\begin{equation}
\Psi_s(u) = \frac{\Phi_s(u)}{u} e^{-i\omega r_*(u)} 
\end{equation}

\noindent
to (\ref{eq:schrodinger}), giving

\begin{eqnarray}
& & u^3 (1-u) g(u)\Phi_s^{\prime\prime}(u)
- u\left[ \frac{u}{2} \left[ 1 - g(u) - (1-3g(u))u \right] 
-4m \sqrt{g(u)}\ i\omega \right]\Phi_s^\prime(u)\nonumber\\
& & + \left[ \frac{u}{2} \left[1 - g(u) - (1-3g(u))u\right]
-\frac{(2m)^2 V(r(u))}{u(1-u)}\right.
\left. -4m\sqrt{g(u)}\ i\omega \right] \Phi_s(u) = 0\, .
\label{eq:pseudo}
\end{eqnarray}

\noindent
where primes represent derivatives with respect to $u$ and 

\begin{equation}
g(u) = 1 -\frac{r_0}{2m}u\, .
\end{equation}

\noindent
Equation~(\ref{eq:pseudo}) would be the same result if we had started
by writing the metric in Eddington-Filkenstein
coordinates~\cite{Mamani:2022akq}. 
The Schwarzschild result is reproduced as $r_0\to 0$.

We now need to solve the above 2'nd-order ordinary differential
equation (ODE)~(\ref{eq:pseudo}) with the
proper boundary conditions.
Substituting the following ansatz for the asymptotic solution

\begin{equation}
\Phi_s(u) = e^{4mi\omega/u} u^{-2(2m+r_0/2)i\omega} \phi_s(u)
\end{equation}

\noindent
into (\ref{eq:pseudo}) gives

\begin{eqnarray}
& & - (1-u) u^3 g\phi_s^{\prime\prime}(u)\nonumber\\
& & +\ \left\{ \frac{1}{2} u^2\left[ 1-g - (1-3g)u \right]
- 4 i u \left[ \sqrt{g} - (3-g)g + (1-g)gu
+ 2gu^2\right) \lambda
  \right\} \phi_s^\prime(u)\nonumber\\ 
& & + \left\{ -\frac{1}{2}u \left[ 1-g - (1-3g)u \right] +
  \frac{(2m)^2 V}{u(1-u)} \right.\nonumber\\ 
& & + i \left[ -3 +4\sqrt{g} - 6g + g^2 + (1-g)^2u + 2(1-g)u^2 \right]
  \lambda\nonumber\\ 
& & \left. + 4\left[ (-6\sqrt{g}+2g^{3/2}+9g-6g^2+g^3)/u -
  4g(2-g)u\right.\right.
\left.    -4gu^2 \right]\lambda^2 \Biggr\} \phi_s(u) = 0\, ,
\label{eq:pseudo2}
\end{eqnarray}

\noindent
where $\lambda = \omega m$.

This is a quadratic eigenvalue problem in $\lambda$ (not the previous
polymerization parameter).
We have solved it using the pseudo-spectral code of
Jansen~\cite{Jansen:2017oag}. 
The following is a brief description of what is implemented in the
code and follows \cite{Mamani:2022akq}.

A quadratic eigenvalue equation can be written as

\begin{equation}
c_0(u,\lambda,\lambda^2)\phi(u) +
c_1(u,\lambda,\lambda^2)\phi^\prime(u) +
c_2(u,\lambda,\lambda^2)\phi^{\prime\prime}(u) = 0\, . 
\end{equation}

\noindent
The coefficients can be written as $c_j(u,\lambda,\lambda^2) =
c_{j,0}(u) + \lambda c_{j,1}(u) + \lambda^2 c_{j,2}(u)$.
The regular function $\phi(u)$ is decomposed into cardinal
functions $C_j(u)$,

\begin{equation}
\phi(u) = \sum_{j=1}^N y(u_j) C_j(u)\, ,
\end{equation}

\noindent
where $y(u)$ is a function of $u$.
The differential equation and functions are evaluated at a set of
points (grid)

\begin{equation}
u_i = \frac{1}{2} \left[ 1 \pm \cos\left(\frac{i\pi}{N}\right)
  \right]\, , \qquad i=0,1,2,\ldots,N\, ,
\end{equation}

\noindent
where it has been written to map the interval $[-1,1]$ into $[0,1]$.
Evaluation on a grid of collection points gives the pseudo-spectral
method its name.
This set of collection points is called the Gauss-Lobatoo grid. 

Since the solution is not periodic and the domain is rectangular,
combinations of Chebyshev polynomials of the first kind are the
cardinal functions of choice:

\begin{equation}
C_j(u) = \frac{2}{N p_j} \sum_{m=0}^N \frac{1}{p_m} T_m(u_j) T_m(u)\,
,\qquad p_0 = p_N = 2, p_j = 1\, . 
\end{equation}

Evaluated on a grid, the coefficients $c_{j,0}(u_i), c_{j,1}(u_i)$,
and $c_{j,2}(u_i)$ can be used to form a matrix:

\begin{equation}
\left( \tilde{M}_0 + \tilde{M}_1\lambda + \tilde{M}_2\lambda^2 \right)
y = 0\, ,
\end{equation}

\noindent
where

\begin{eqnarray}
(\tilde{M}_0)_{ji} & = c_{0,0}(u_i) D_{ji} + c_{1,0}(u_i)
  D_{ji}^{(1)} + c_{2,0}(u_i) D_{ji}^{(2)}\, ,\\ 
(\tilde{M}_1)_{ji} & = c_{0,1}(u_i) D_{ji} + c_{1,1}(u_i)
  D_{ji}^{(1)} + c_{2,1}(u_i) D_{ji}^{(2)}\, ,\\ 
(\tilde{M}_2)_{ji} & = c_{0,2}(u_i) D_{ji} + c_{1,2}(u_i)
  D_{ji}^{(1)} + c_{2,2}(u_i) D_{ji}^{(2)}\, ,
\end{eqnarray}

\noindent
and $D_{ji}, D_{ji}^{(1)}, D_{ji}^{(2)}$ are matrices of the cardinal
function and its derivatives. 
Defining $\tilde{y} = \lambda y$,

\begin{equation}
\tilde{M}_0 y + \left( \tilde{M}_1 + \tilde{M}_2\lambda \right)
\tilde{y} = 0\, . 
\end{equation}

\noindent
The matrix representation of the eigenvalue problem becomes

\begin{equation}
(M_0 + M_1\lambda)\cdot \vec{y} = 0\, ,
\end{equation}

\noindent
where

\begin{equation}
  M_0 = \left(
  \begin{array}{cc}
  \tilde{M}_0 & \tilde{M}_1\\ 0 & 1
  \end{array}
  \right)\, , \quad 
  M_1 = \left(
  \begin{array}{cc}
  0 & \tilde{M}_2\\ -1 & 0
  \end{array}
  \right)\, , \quad \textrm{and} \quad
  \vec{y}= \left(
  \begin{array}{c}
  y\\ \tilde{y}
  \end{array}
  \right)\, .
\end{equation}

Calculating QNM frequencies using the PSM does not depend on any
initial guess, as in other methods.
Unfortunately, the PSM leads to spurious solutions that do not have
any physical meaning. 
The problem then becomes one of detecting and eliminating spurious
solutions.

We have rejected spurious frequencies by repeating the calculations
using different grid sizes and precisions, and selecting common
frequencies in the two cases.
We also found it very effective to remove frequencies with extremely
small real components, as often happens in this method.
The latter is valid since the PSM is not probing high enough overtones with
vanishing real components.

\subsection{Continued fraction method}

The CFM was first used to solve for QNMs by
Leaver~\cite{Leaver:1985ax}. 
The method essential solves the differential equation by Frobenius
method resulting in a set of algebraic recurrence relations, which
are solved by continued fractions.

The \SH-like wave equation~(\ref{eq:schrodinger}) is simple in $r_*$ and
is good for applying the boundary condition.
However, because of the potential, it can be better to solve the
equation in the variable $r$. 
The 2'nd-order ODE in $r$ becomes

\begin{eqnarray}
& & (r-2m)(r-r_0)\Psi_s^{\prime\prime}(r) + \frac{2m(r-r_0)
  + (r_0/2)(r-2m)}{r} \Psi_s^\prime(r)\nonumber\\
& & + \frac{r^3}{r-2m} [\omega^2 - V_s(r)]\Psi_s(r) = 0\, ,
\label{eq:ode1}
\end{eqnarray}

\noindent
which can be written in general as

\begin{equation}
P(r) \Psi_s^{\prime\prime} + Q(r)\Psi_s^\prime + R(r)\Psi_s = 0\, ,
\label{eq:ode2}
\end{equation}

\noindent
where

\begin{eqnarray}
  P & = & (r-2m)(r-r_0)\, ,\\
  Q & = & \frac{2m(r-r_0) + (r_0/2)(r-2m)}{r}\, ,\\
  R & = & \frac{r^3}{r-2m} (\omega^2 - V_s(r))\, .
\end{eqnarray}

We postulate a series solution consisting of the asymptotic wave
function $\psi(r)$, and the series approximation $\phi_s(r)$: $\Psi(r)
= \psi(r) \phi_s(r)$.  
Applying the chain rule for $\Psi$ and substitution into
(\ref{eq:ode2}) gives

\begin{equation}
P\psi\phi_s^{\prime\prime} + (2P\psi^\prime+Q\psi)\phi_s^\prime +
(P\psi^{\prime\prime}+Q\psi^\prime+R\psi)\phi_s = 0\, . \label{eq:ode3} 
\end{equation}

For the asymptotic solution, we take the form written in
\cite{Moreira:2023cxy}, which simplifies the recurrence relations
relative to using the form presented in appendix~\ref{app:B}:  

\begin{equation}
\psi(r) = r(r-r_0)^{i\omega(2m+r_0/2)-1} e^{i\omega r}\, .
\end{equation}

\noindent
The difference amounts to replacing $r$ by $r-r_0$ for some powers,
which is allowed as $r\to\infty$.

In (\ref{eq:ode1}), regular singularities occur at $r=0$, $r=r_0$,
and $r=2m$, while $r\to\infty$ is an irregular singularity. 
Following \cite{Moreira:2023cxy}, the singularities at
$(0,r_0,2m,\infty)$ are mapped to $(2m/r_0,\infty,0,1)$ by the
change of variable

\begin{equation}
  u = \frac{r-2m}{r-r_0}\, .
\end{equation}

\noindent
The domain $[2m,\infty]$, now maps to $[0,1]$ which is advantageous
for numerical computation.

The variable $u$ is also the expansion variable in the Frobenius
series. 
The derivatives of the series are ($r_0\ne 2m$) 

\begin{eqnarray}
\phi_s(u) & = & \sum_{n=0}^\infty a_n u^{\zeta+n}\, ,\\
\phi_s^\prime(u) & = & \frac{(1-u)^2}{2m-r_0} \sum_{n=0}^\infty a_n
(\zeta+n) u^{\zeta+n-1}\, ,\\ 
\phi_s^{\prime\prime}(u) & = & \frac{(1-u)^3}{(2m-r_0)^2}
\sum_{n=0}^\infty a_n (\zeta+n) \left[ (\zeta+n-1)(1-u)
  u^{\zeta+n-2}\right. 
\left. - 2 u^{\zeta+n-1} \right]\, ,
\end{eqnarray}

\noindent
where primes now represent differentiation with respect to $r$.
The characteristic exponent $\zeta$ is determined by the indicial
equation and is found to be

\begin{equation}
\zeta = \frac{-i\omega (2m)^{3/2}}{\sqrt{2m-r_0}}\, .
\end{equation}

\noindent
The indicial equation corresponds to the ingoing solution at the
horizon (see appendix~\ref{app:B}).

For $s=0,1,2$ (axial), we obtain upon substitution of the series
solution, the following four-term recurrence relations  

\begin{eqnarray}
\alpha_0 a_1 + \beta_0 a_0 & = & 0\, ,\\
\alpha_1 a_2 + \beta_1 a_1 + \gamma_1 a_0 & = & 0\, ,\\
\alpha_n a_{n+1} + \beta_n a_n + \gamma_n a_{n-1} + \delta_n a_{n-2} &
= & 0, \qquad n = 2, 3, \ldots\, ,
\end{eqnarray}

\noindent
where

\begin{eqnarray}
\alpha_n & = & -8(n+1) i\omega (2m)^{5/2} (2m-r_0)
            + 4 (n+1)^2 (2m) (2m-r_0)^{3/2}\, ,\\ 
\beta_n  & = & 4 \omega^2 (2m)^4 (2m-r_0)^{1/2}\nonumber\\
          && + 2 [(12n+5) i\omega + 6 \omega^2 (2m)] (2m)^{5/2}
            (2m-r_0)\nonumber\\
          && - 2 [(6n^2+5n+2) + 2[\ell(\ell+1) - s(s-1)]\nonumber\\
          && - 3(2n+1) i\omega (2m) - 6\omega^2 (2m)^2] (2m)
            (2m-r_0)^{3/2}\nonumber\\ 
          && - 2 [(4n+1) i\omega - 2 \omega^2 (2m)] (2m)^{3/2}
            (2m-r_0)^2\nonumber\\  
          && + 2[2s + (2n+1)(n+ i\omega (2m))] (2m-r_0)^{5/2}\, ,\\
\gamma_n & = & -8 \omega^2 (2m)^4 (2m-r_0)^{1/2}\nonumber\\
          && - 4 [(6n-1) i\omega + 6\omega^2 (2m)] (2m)^{5/2}
            (2m-r_0)\nonumber\\
          && + [ 2(6n^2-2n+1) + 4 [\ell(\ell+1)-s(s-1)]\nonumber\\
          && - (24n-3) i\omega (2m) - 17 \omega^2 (2m)^2] (2m)
            (2m-r_0)^{3/2}\nonumber\\ 
          && + 4 [(4n-1) i\omega + 3 \omega^2 (2m)](2m)^{3/2}
            (2m-r_0)^2\nonumber\\ 
          && - 2 [(4n^2-2n+1) + 2 \ell(\ell+1) -s(2s-1)\nonumber\\
          && - (6n-1) i\omega (2m) - 9\omega^2 (2m)^2]
            (2m-r_0)^{5/2}\nonumber\\    
          && + 4 \omega^2 (2m)^{3/2} (2m-r_0)^3\nonumber\\ 
          && - [2s(2s+1)/(2m) - (4n-1) i\omega + \omega^2 (2m)]
            (2m-r_0)^{7/2}\, ,\\ 
\delta_n & = & 4 \omega^2 (2m)^4 (2m-r_0)^{1/2}
            + 2 [(4n-3) i\omega + 6 \omega^2 (2m)] (2m)^{5/2}
            (2m-r_0)\nonumber\\
          && - 2 [2(2n^2-3n+1) - (12n-9) i\omega (2m)\nonumber\\
          && - 5\omega^2 (2m)^2) (2m) (2m-r_0)^{3/2}\nonumber\\ 
          && - [(4n-3) i\omega + 8 \omega^2 (2m)] (2m)^{3/2}
            (2m-r_0)^2\nonumber\\ 
          && + [2(2n^2-3n+1) - 4(4n-3) i\omega (2m) - 15 \omega^2
            (2m)^2] (2m-r_0)^{5/2}\nonumber\\ 
          && + 4 \omega^2 (2m)^{3/2} (2m-r_0)^3\nonumber\\
          && - [2s(2s+1)/(2m) - (4n-3)i\omega - 7 \omega^2 (2m)]
            (2m-r_0)^{7/2}\nonumber\\
          && + [2s(2s+1)/(2m)^2 - \omega^2] (2m-r_0)^{9/2}\, ,
\end{eqnarray}

\noindent 
The result agrees with \cite{Moreira:2023cxy} for $s=0$.
Notice that all coefficients vanish as $r_0\to 2m$.
The coefficients are at most quadratic in $\omega$ and the terms have
dimension of mass to the 5/2 power.
The $\sqrt{2m-r_0}$ factor comes from the inducial index.

The four-term recurrence relation is reduced to a three-term
recurrence relation by applying Gaussian
elimination~\cite{PhysRevD.41.2986}:

\begin{equation}
  \tilde{\alpha}_n = \alpha_n\, , \quad
  \tilde{\beta}_n = \beta_n\, ,   \quad
  \tilde{\gamma}_n = \gamma_n\, , \quad \textrm{and} \quad
  \tilde{\delta}_n = 0\, , \qquad \textrm{for}\ n=0,1\, ,
\end{equation}

\noindent
and

\begin{equation}
  \tilde{\alpha}_n = \alpha_n\, , \quad
  \tilde{\beta}_n = \beta_n -
  \frac{\tilde{\alpha}_{n-1}\delta_n}{\tilde{\gamma}_{n-1}}\, , \quad
  \textrm{and} \quad
  \tilde{\gamma}_n = \gamma_n -
  \frac{\tilde{\beta}_{n-1}\delta_n}{\tilde{\gamma}_{n-1}}\, ,
  \quad \textrm{for}\ n\ge 2\, .  \label{eq:tdef}
\end{equation}

\noindent
The new variables now satisfies the three-term recurrence relation

\begin{eqnarray}\label{eq:reca}
\tilde{\alpha}_0 a_1 + \tilde{\beta}_0 a_0 & = & 0\, ,\\
\tilde{\alpha}_n a_{n+1} + \tilde{\beta}_n a_n + \tilde{\gamma}_n
a_{n-1} & = & 0, \qquad n = 1, 2, \ldots\, .
\label{eq:recb}
\end{eqnarray}

The recurrence relations can be solved using a continued fraction.
Writing (\ref{eq:recb}) as

\begin{equation}
\frac{a_n}{a_{n-1}} =
\frac{-\tilde{\gamma}_n}{\tilde{\beta}_n+\tilde{\alpha}_n 
  a_{n+1}/a_n}\, , 
\end{equation}

\noindent
we substitution the left side into the right side an infinite number of
times and use (\ref{eq:reca}) to give the continued fraction

\begin{equation}
0 = \tilde{\beta}_0 -
\frac{\tilde{\alpha}_0\tilde{\gamma}_1}{\tilde{\beta}_1-} 
\frac{\tilde{\alpha}_1\tilde{\gamma}_2}{\tilde{\beta}_2-}
\frac{\tilde{\alpha}_2\tilde{\gamma}_3}{\tilde{\beta}_3-} \cdots\, .
\end{equation}

The $n$'th inversion of the continued fraction is

\begin{equation}
  \tilde{\beta}_n -
  \frac{\tilde{\alpha}_{n-1}\tilde{\gamma}_n}{\tilde{\beta}_{n-1}-}
  \cdots - 
  \frac{\tilde{\alpha}_0\tilde{\gamma}_1}{\tilde{\beta}_0} =
  \frac{\tilde{\alpha}_n\tilde{\gamma}_{n+1}}{\tilde{\beta}_{n+1}-}
  \frac{\tilde{\alpha}_{n+1}\tilde{\gamma}_{n+2}}{\tilde{\beta}_{n+2}-}
  \cdots   
\end{equation}

\noindent
According to \cite{Leaver:1985ax}, the $n$'th inversion is the
formula that should be used to solve for the $n$'th eigenfrequency. 
The left side of the $n$'th inversion is a finite continued
fraction and is calculated using ``back calculation'' starting with
$\tilde{\alpha}_0\tilde{\gamma}_1/\tilde{\beta}_0$. 
We have calculated the right side of the $n$'th inversion using the
Nollert's remainder.

Nollert~\cite{PhysRevD.47.5253} devised a method of improve on
Leaver's continued fraction method by estimating the remainder when
truncating the back calculation.
To get the Nollert method to work with four-term recurrence relations
and Gaussian elimination, a large $n$ approximation of the tilde
coefficients can be calculated.

We first obtain large $n$ asymptotic limits of the $\alpha_n,
\beta_n, \gamma_n$, and $\delta_n$ coefficients by dividing their
expressions by $n^2$ and keeping the first two terms in the expansion.
Next we postulate the large-$n$ asymptotic expansion of
$\tilde{\alpha}_n, \tilde{\beta}_n$, and $\tilde{\gamma}_n$ to be of
the form (also to 2'nd-order) 

\begin{eqnarray}
\tilde{\alpha}_n & = r + u/n\, ,\\
\tilde{\beta}_n  & = s + v/n\, ,\\
\tilde{\gamma}_n & = t + w/n\, .
\end{eqnarray}

\noindent
We then substitute the large-$n$ coefficients into the definitions
(\ref{eq:tdef}) to obtain

\begin{eqnarray}
\tilde{\alpha}_n & = & 8 m (2m-r_0)^{3/2} + u/n\, ,\\
\tilde{\beta}_n  & = & -16 m (2m-r_0)^{3/2} + v/n\, ,\\
\tilde{\gamma}_n & = & 8 m (2m-r_0)^{3/2} + w/n\, ,
\end{eqnarray}

\noindent
where

\begin{eqnarray}
u & = & 8[-i\omega(2m)^{5/2}(2m-r_0) + (2m)(2m-r_0)^{3/2}]\, ,\\
v & = & 4[4i\omega(2m)^{5/2}(2m-r_0) -
  2[1-3i\omega(2m)](2m)(2m-r_0)^{3/2}\nonumber\\
  & & + i\omega(2m)(2m-r_0)^{5/2}]\, ,\\ 
w & = & -4i\omega[2(2m)^{5/2}(2m-r_0) + 3(2m)^2(2m-r_0)^{3/2} - (2m)(2m-r_0)^{5/2}]\, .
\end{eqnarray}

\noindent
The recurrence relationship for the Nollert reminder is

\begin{equation}
R_N = \frac{\tilde{\gamma}_N}{\tilde{\beta}_N - \tilde{\alpha}_N
  R_{N+1}}\, ,
\end{equation}

\noindent
which can be expanded as

\begin{equation}
R_N = C_0 + C_1 N^{-1/2} + C_2 N^{-1} + \cdots\, .
\end{equation}

\noindent
The Nollert coefficients are

\begin{eqnarray}
C_0 & = & -1\, ,\\
C_1 & = & \pm\sqrt{2i(2m-r_0)\omega}\, ,\\\label{eq:nolert}
C_2 & = & \frac{3}{4} - 2i\left(2m-\frac{r_0}{4}\right)\omega\, ,
\end{eqnarray}

\noindent
where the sign is chosen such that $\mathcal{R}e [C_1] > 0$.

We have validated our Nollert method by also using a straight
back-calculation method and the modified Lentz's
method~\cite{Lentz:76}.  
In the limit of vary large number of fractions, back calculation
and Nollert's method should have the same accuracy.
In the back-calculation method, for some large value of $N$ we take
$\tilde{\alpha}_N = 0$.
This allows a backward calculation of the right side of the $n$'th
inversion back to the fraction containing
$\tilde{\alpha}_n\tilde{\gamma}_{n+1}/\tilde{\beta}_{n+1}$. 
A good approximation method for evaluating continued fractions is the
modified Lentz's method~\cite{Lentz:76}.
It is based on relating continued fractions to rational
approximations, and allows a simple test of how much the result
changes from one iteration to the next.

For any of the three methods, the resulting equation in $\omega$ must
be solved by root-finding.  
A guess for the root is obtained by linear extrapolation using the
current $\omega_n$ and the change given by the difference between the
current $\omega_n$ and the previous $\omega_{n-1}$.
An approximation to the $n=0$ mode is used as the initial guess.

%
\section{Results\label{sec:results}}

We present QNMs on a phase diagram versus the quantum parameter $r_0$,
higher overtones for different values of $r_0$, and isospectrality
violation by examining the difference in QNMs of axial- and
polar-gravitational perturbations. 
When referring to the Schwarzschild value, $r_0=0$ is used in the ABV
calculations.

\subsection{Trajectories}

\begin{figure*}[htb]
\includegraphics[width=0.49\linewidth]{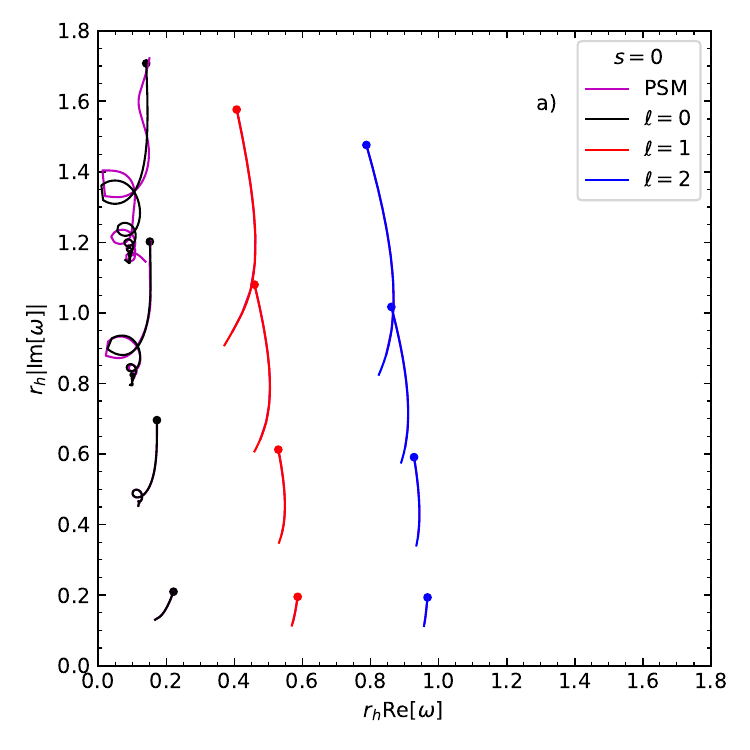}
\includegraphics[width=0.49\linewidth]{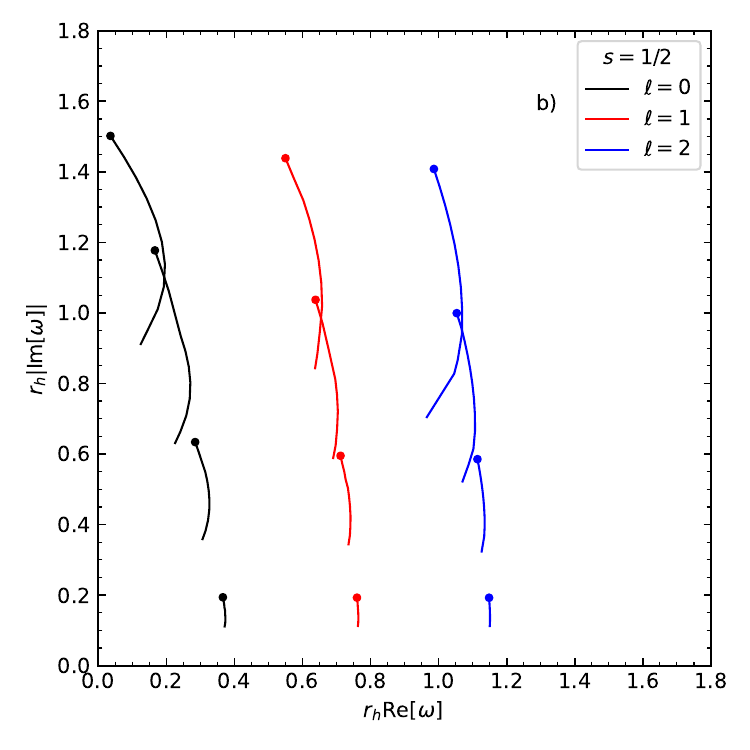}
\includegraphics[width=0.49\linewidth]{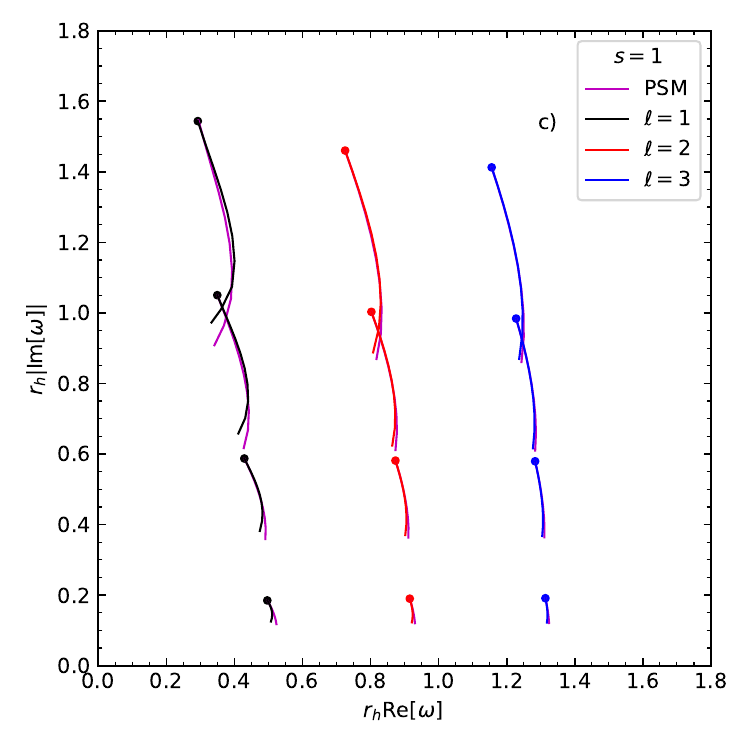}
\includegraphics[width=0.49\linewidth]{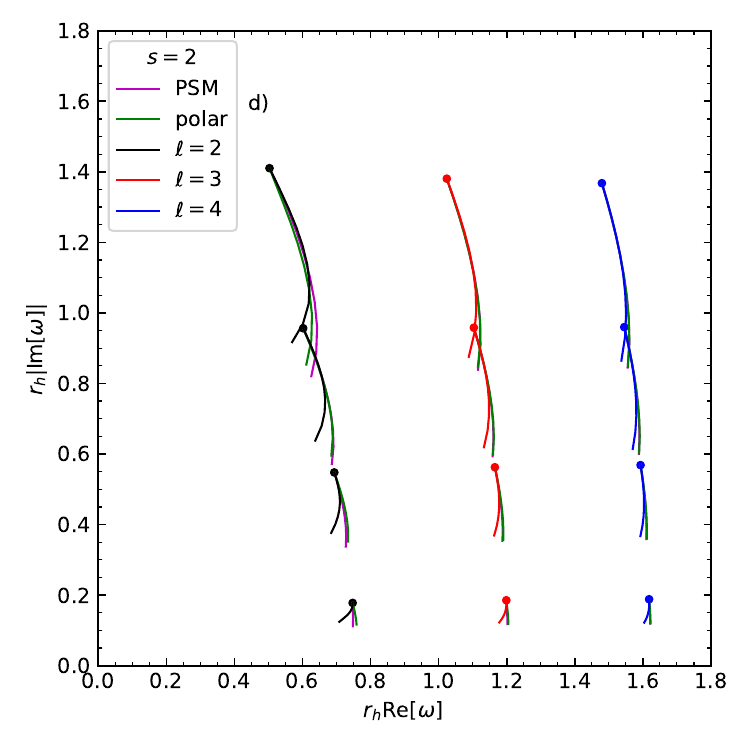}
\caption{\label{fig:traj}First four ($n=0,1,2,3$) QNMs from bottom to
  top for $r_0 = 0$ to 0.9 (0.99 for spirals).
  a) $s=0$   and $\ell=0,1,2$ from left to right,
  b) $s=1/2$ and $\ell=0,1,2$ from left to right,
  c) $s=1$   and $\ell=1,2,3$ from left to right, and
  d) $s=2$   and $\ell=2,3,4$ from left to right.
  The top points of each curve are the Schwarzschild values.
  A mass of $m=1/2$ has been used.}
\end{figure*}

We first study how the QNMs change for different values of $r_0$.
Figure~\ref{fig:traj} shows phase diagrams or trajectories for
$r_0 = [0,0.99]$ for $s=\ell=0$ and $n=1,2,3$, and $r_0 = [0,0.9]$
otherwise. 
The CFM and PSM were used for $s=0, 1, 2$ (axial), and WKB method for
$s=1/2$.
The PSM was also used for $s=2$ (polar).
The CFM calculations are shown in black, red, and blue.
Underneath these curves the PSM calculations are shown in magenta.
In some cases the CFM and PSM are similar enough that the PSM curves
are not visible.
The difference between the two method of calculation are most apparent
for large $r_0$. 
In the $s=2$ plot, the PSM calculations of the polar modes are shown
underneath in green. 

For $s=0$, we reproduce the results of \cite{Moreira:2023cxy} and
continue them to overtone $n=3$.
The $s=1$ and $s=2$ results are new, although for $s=1$
\cite{Fu:2023drp} plotted the real and imaginary parts of $\omega$
separately.
The self-intersecting spirals continue for $s=\ell=0$.
We do not observe any additional self-intersecting spirals beyond
those.

The trajectories using two different numerical methods agree well for
$s=0$ in the non-spiral cases.
The agreement is less good for $s=1$ and $s=2$.
However, the agreement improves as $\ell$ increases.
It's clear that the CFM determines the self-intersecting spirals better
than the PSM. 
The $s=1/2$ trajectories are unlikely to be particularly accurate for
$n > \ell$.

Self-intersecting and non-intersecting spiral trajectories have been
previously observed for Reissner-Nordstr\"{o}m and Kerr black
holes~\cite{PhysRevD.41.2986,Jing:2008an,Berti:2003zu}, as well as
modified gravity black holes~\cite{Konoplya:2022hll,Zhang:2024nny}.  

\subsection{Higher overtones}

\begin{figure*}[htb]
\includegraphics[width=0.49\linewidth]{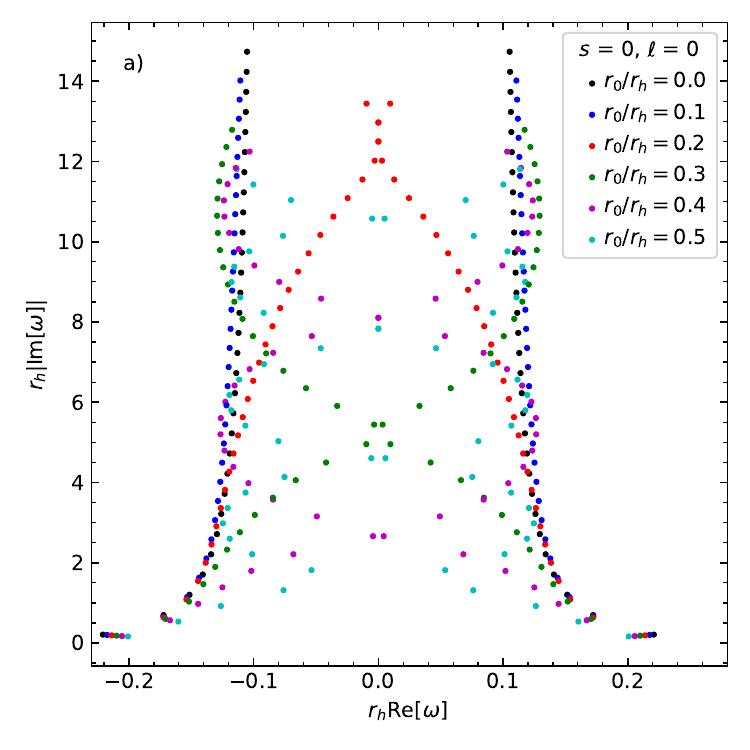}
\includegraphics[width=0.49\linewidth]{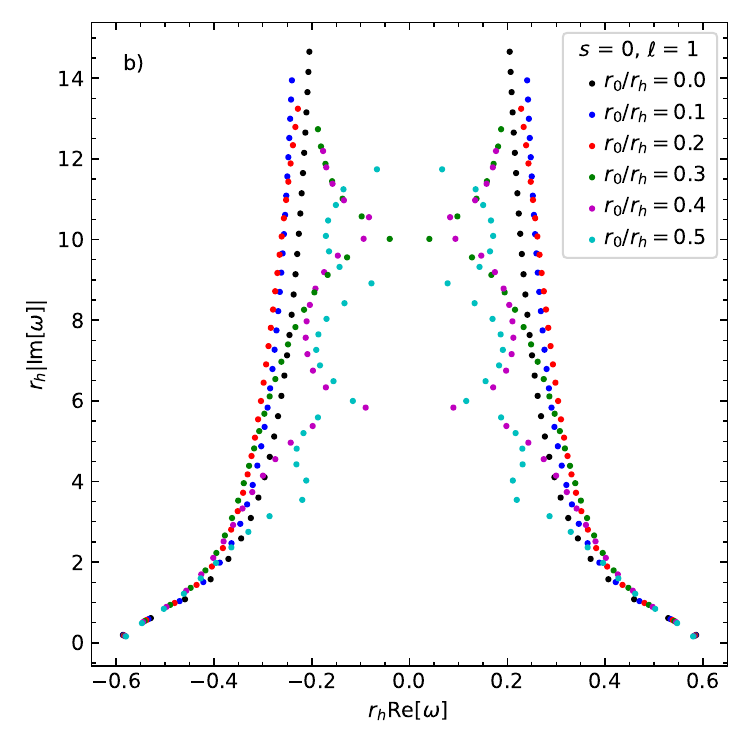}
\caption{\label{fig:tones0}First 30 QNMs $(n=0,1,\ldots 29)$ for
  scalar perturbations on a holonomy corrected black hole background
  a) $s=0, \ell=0$ and b) $s=0, \ell=1$,
  The continued fraction method with mass $m=1/2$ was used.} 
\end{figure*}

\begin{figure*}[htb]
\includegraphics[width=0.49\linewidth]{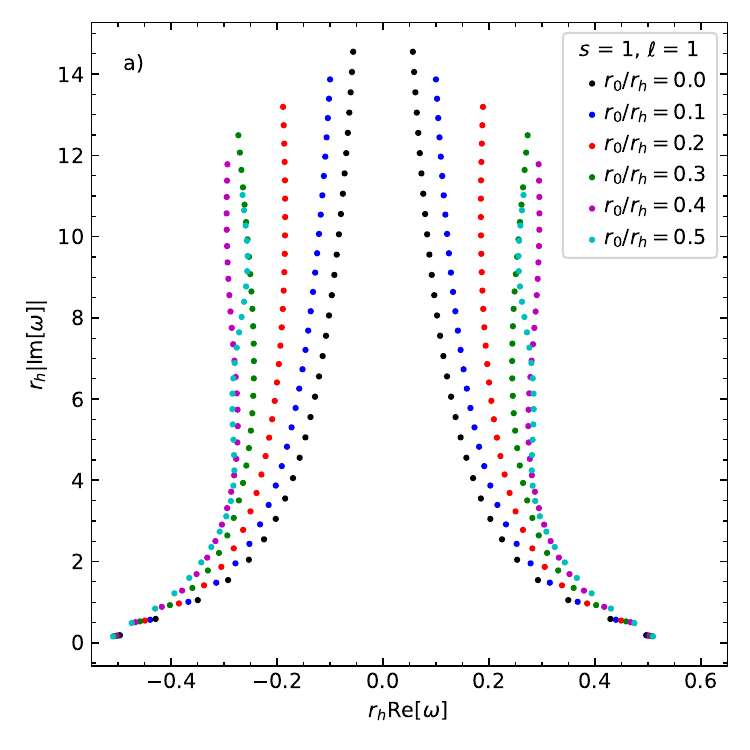}
\includegraphics[width=0.49\linewidth]{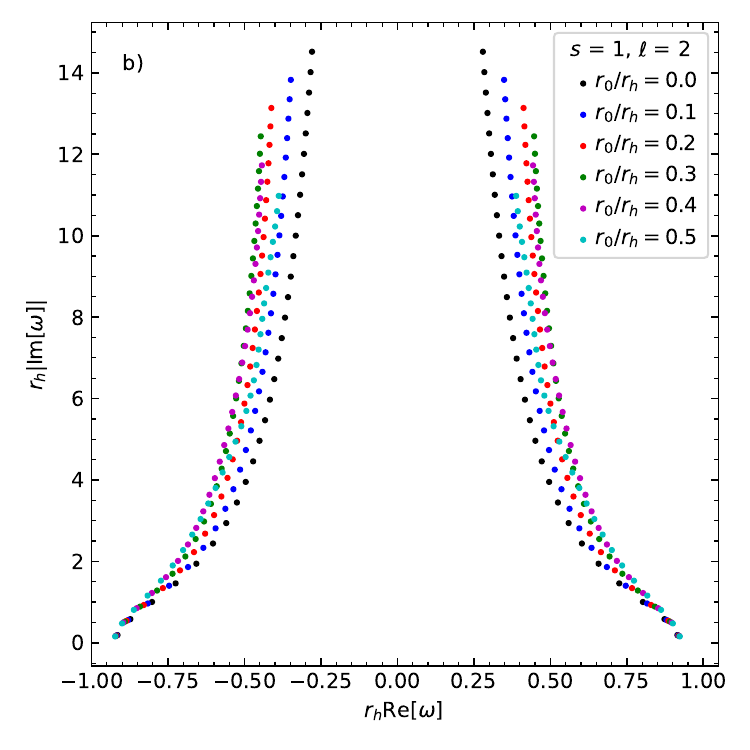}
\caption{\label{fig:tones1}First 30 QNMs $(n=0,1,\ldots 29)$ for
  vector perturbations on a holonomy corrected black hole background
  a) $s=1, \ell=1$ and b) $s=1, s=2$.
  The continued fraction method with mass $m=1/2$ was used.} 
\end{figure*}

\begin{figure*}[htb]
\includegraphics[width=0.49\linewidth]{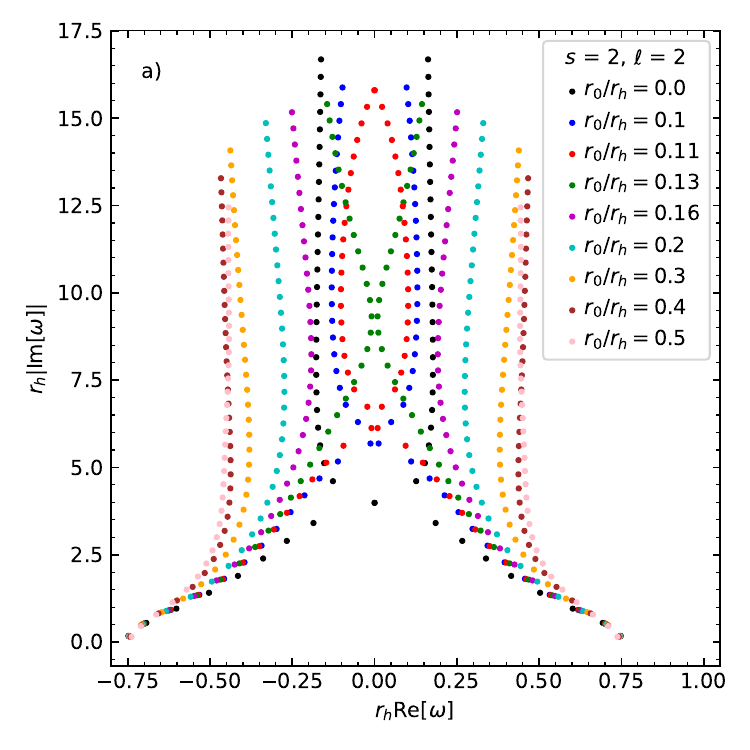}
\includegraphics[width=0.49\linewidth]{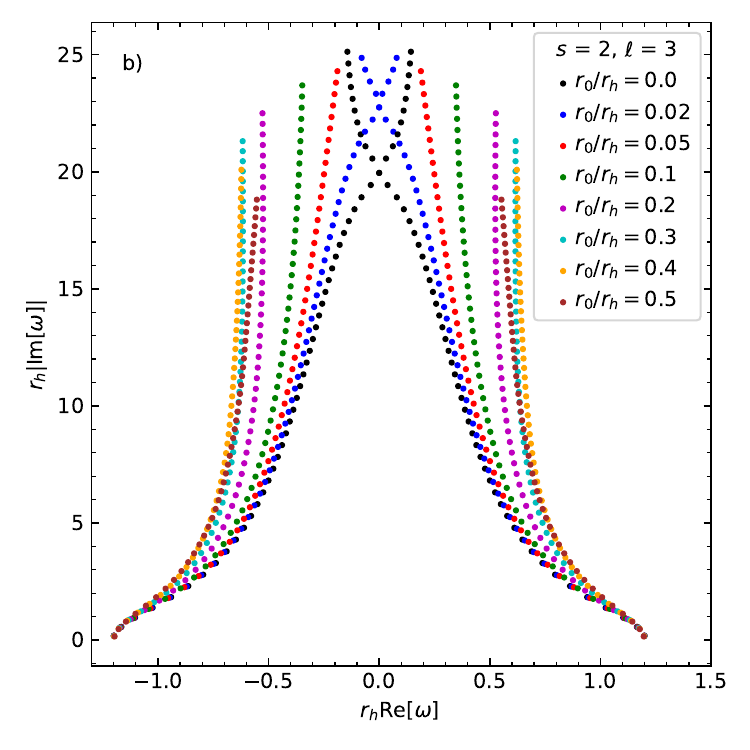}
\caption{\label{fig:tones2}Higher overtones for
  axial-gravitational perturbations on a holonomy corrected black
  hole background
  a) first 34 QNMs $(n=0,1,\ldots 33)$, $s=2$ (axial), $\ell=2$ and
  b) first 51 QNMS $(n=0,1,\ldots 50)$, $s=2$ (axial), $\ell=3$.
  The continued fraction method with mass $m=1/2$ was used.} 
\end{figure*}

The ABV metric is very close to Schwarzschild everywhere, except for a
small region near the event horizon, which is crucial for
overtones~\cite{Konoplya:2022pbc}. 
Figures~\ref{fig:tones0}, \ref{fig:tones1}, and \ref{fig:tones2} show
higher overtones using the CFM.
Based on the trajectory agreement between the PSM and CFM for low
values of $r_0$, we only consider $r_0 \le 0.5$ when calculating the
higher overtones.  
We have assumed symmetry in the real part of $\omega$ and drawn
identical points on both sides of the imaginary axis.

For $s=0$ (figure~\ref{fig:tones0}), we reproduce the results in
\cite{Moreira:2023cxy}.
For $s=1$ (figure~\ref{fig:tones1}) and $s=2$ (axial)
(figure~\ref{fig:tones2}), the results are new. 
For $s=1$, within the number of modes considered, no oscillations or
pure imaginary modes are observed.

Interesting behaviour is observed for $s=2$ (figure~\ref{fig:tones2}).
Purely imaginary modes occur above the, well known, Schwarzschild
purely imaginary modes for small values of $r_0$. 
It seems likely that pure imaginary modes will occur for other
values of $r_0$. 
As $r_0$ increases the trend is similar to the $s=1$
perturbations, and it is not clear if purely imaginary modes will
occur at very high overtones.
No multiple oscillations are observed for $s=2$ over the number of
overtones we calculate.


\subsection{Isospectrality}

If the QNM frequency spectrum from axial- and polar-gravitational
perturbations are the same, this property is referred to as
isospectrality. 
Isospectrality for Schwarzschild and Reissner-Nordstr{\"o}m metrics
have been proven~\cite{Chandrasekhar,chandrasekharRN}. 
Isospectrality has also been demonstrated to linear order in the spin
for Kerr black holes~\cite{Pani:2013ija,Pani:2013wsa}, and for
Schwarzschild-de Sitter and Schwarzschild-anti de Sitter
spacetimes~\cite{Moulin:2019bfh}. 
However, isospectrality appears not to be a universal feature that
holds for modified gravity spacetimes~\cite{
Prasobh:2014zea,
Bhattacharyya:2017tyc,
Bhattacharyya:2018hsj,
Cruz:2020emz,
Chen:2021pxd,
del-Corral:2022kbk}.
It is thus important to compare the spectra for gravitational
perturbations on the ABV black hole background.

\begin{figure*}[htb]
\includegraphics[width=0.49\linewidth]{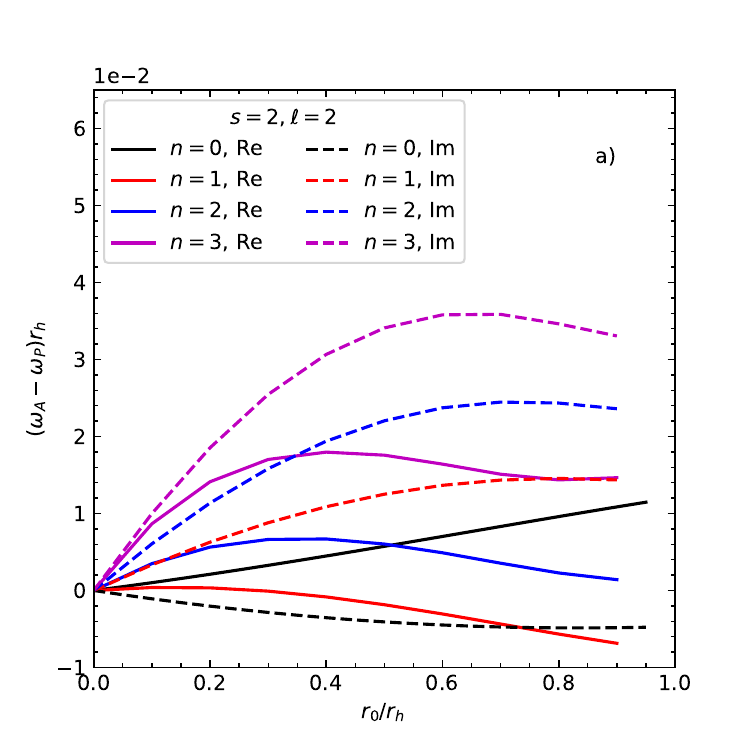}
\includegraphics[width=0.49\linewidth]{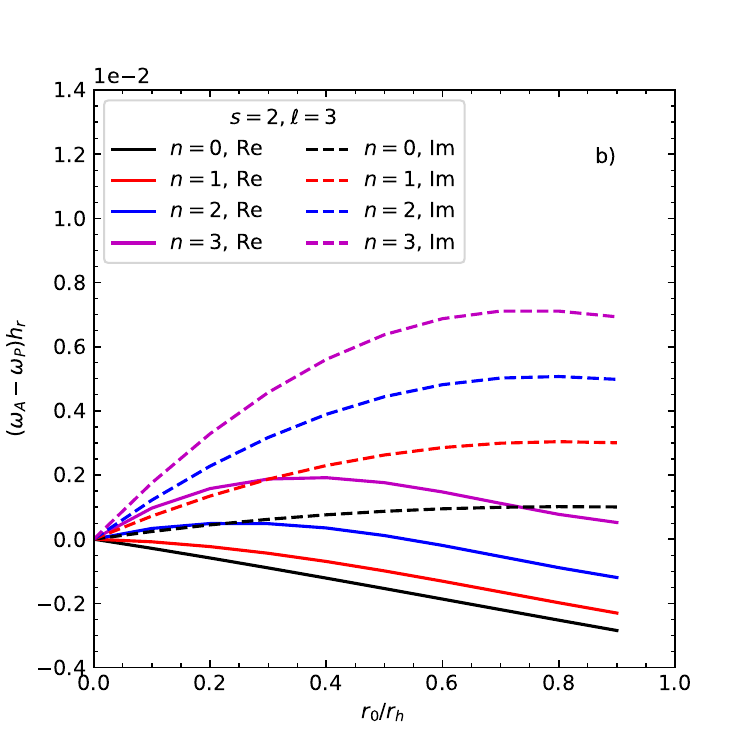}
\caption{\label{fig:iso}Difference between the axial- and
  polar-gravitational quasinormal mode frequencies versus $r_0$ on a  
  holonomy corrected black hole background. The difference in the real
  and imaginary parts are shown as solid and dashed lines,
  respectively. The pseudo-spectral method with mass of $m=1/2$ has
  been used.}   
\end{figure*}

Figure~\ref{fig:iso} shows the difference between the axial $\omega_A$
and polar $\omega_P$ gravitational quasinormal mode frequencies versus
$r_0$ on the ABV black hole background.
The difference in the real and imaginary parts are shown as solid and
dashed lines, respectively.
Schwarzschild isospectrality was obtained to better than seven
digits.
The maximum violation of isospectrality is observed to not necessarily
occur at the extremal value of $r_0=r_h$.
The amount of isospectrality violation, in general, increases with
overtone number.
The case of fundamental mode $n=0$ for $\ell=2$ shows nonmonotonic
variation in the real part relative to the overtones and the
fundamental for higher values of $\ell$.
The amount of isospectrality violation decreases with increasing $\ell$.
The violation of isospectrality is about a factor of 0.3 for $\ell=3$
relative to $\ell=2$, and then again a factor of about 0.5 for
$\ell=4$ (not shown) relative to  $\ell=3$ .

%
\section{Discussion\label{sec:discussion}}

We have plotted the QNMs versus $r_0$ on a phase-space diagram for
$s=0, 1, 2$ and $\ell = s, s+1, s+2$ for $n=0,1,2,3$.
As well as, $s = 1/2$, for $\ell = 0, 1, 2$.
Only the $s=\ell=0$, $n=1,2,3$ modes exhibit self-intersecting
spirals. 
We postulate that such behaviour is unlikely to be a universal
property of the ABV metric but more associated to the unique shape of
the potentials for $s=0$.
Such behaviour has previously been observed, for example, in
Reissner-Nordstr{\"o}m spacetime~\cite{PhysRevD.41.2986}.
The ABV behaviour may not be unexpected if we interpret the
Reissner-Nordstr{\"o}m mass and charge to be $M = m + r_0/2$ and $Q^2
= 2mr_0$, respectively~\cite{Alonso-Bardaji:2022ear}.

We now discuss the trajectories in terms of the polymerization
approach. 
Recall that $r_0$ given by (\ref{eq:r0}) is proportional to the
constant of the motion  $m$, and $\lambda$ is a real dimensionless
parameter, which can be taken without loss of generality to be
positive. 
It encodes the discretization of the quantum spacetime and is related
to the length of the holonomies. 
We can consider $\lambda$ a universal constant over the entire phase 
space in which case the trajectories will depend on $\lambda$.
Or, we can consider $\lambda$ a constant of the solution, in which the
trajectories will depend on $r_0$.
The trajectory plots support either interpretation, although in the
first interpretation it should be understood that $\lambda$ is varying
along the trajectory, not $r_0$ as mentioned in the figure captions.

Higher overtones for $s=0, 1, 2$ (axial) and for $\ell=s$ and
$\ell=s+1$ are calculated. 
The $s=0$ behaviour first observed in \cite{Moreira:2023cxy} is
reproduced.
For $s=1$, there appears to be no case of near vanishing oscillating
modes.
However, for axial-gravitational perturbations a rich pattern of
overtones is observed for $\ell=2$.
In this case, we are probing beyond the first crossing of the
imaginary axis.
For $\ell=3$, oscillations in the real part of the QNMs may occur for 
very high overtones.

Isospectrality is clearly violated for gravitational perturbations on
a ABV black hole background.
In general, the amount of violation increases with increasing overtone
number and decreases with increasing $\ell$.
The amount of isospectrality violation is not always monotonically
increasing with increasing $r_0$, and is not, in general, a maximum or 
minimum at the extremal value of $r_0 = r_h$.

\acknowledgments
We acknowledge the support of the Natural Sciences and Engineering
Research Council of Canada (NSERC). 
Nous remercions le Conseil de recherches en sciences naturelles et en
g{\'e}nie du Canada (CRSNG) de son soutien. 
\appendix 
\section{Gravitational perturbations\label{app:A}}

In absence of an underlying theory, we assume the ABV effective
quantum black hole is governed by the Einstein equation with an
effective energy-momentum tensor.
We can view it as Einstein gravity minimally coupled to an anisotropic
fluid, where it is the anisotropic fluid that drives the quantum 
corrections.
The quantum-corrected metric in vacuum is applied to the left-hand
side of the Einstein equation to give a non-zero effective
energy-momentum tensor on the right-hand side of the equation.
This effective matter field is expected to violate the energy
conditions.

To study gravitational perturbations, one needs to perturb the
gravitational equation, as well as, the energy-momentum tensor. 
Generally, the gravitational perturbations will affect symmetries of
the background spacetime and the form of the modified Einstein
equation. 

We assume the quantum corrections to Einstein equation are
also of the anisotropic fluid form in the perturbation level.
The equation describing the gravitational perturbations is thus
derived by perturbing the spacetime metric, the energy-momentum
tensor of the anisotropic fluid, and the Einstein equation.

The perturbed spacetime can be described by a non-stationary and
axisymmetric metric in which the symmetrical axis is turned such that
no $\phi$ dependence appears in the metric
functions~\cite{Chandrasekhar}: 

\begin{equation}
ds^2 = -e^{2\nu} dt^2 + e^{2\psi} (d\phi - \sigma dt - q_2
dr - q_3 d\theta)^2 +
e^{2\mu_2} dr^2 + e^{2\mu_3} d\theta^2\, , 
\end{equation}

\noindent
where, in general, $\nu, \psi, \mu_2, \mu_3, \sigma, q_2, q_3$ depend
on the time coordinate $t$, radial coordinate $r$, and polar angle
$\theta$. 
When linearizing the field equations, axial perturbations are encoded
in the function $\sigma, q_2, q_3$, while the other metric functions
$\nu, \psi, \mu_2, \mu_3$ remain zeroth order quantities and are
functions of $r$ only. 
Note that $\sigma, q_2, q_3$ are zero for a static and spherically
symmetric spacetime.

Because of the symmetries of the background spacetime, the effective
energy-momentum tensor of this anisotropic perfect fluid can be
written as

\begin{equation}
T_{\mu\nu} = (\rho + p_2) u_\mu u_\nu + (p_1-p_2) x_\mu x_\nu + p_2
g_{\mu\nu}\, ,
\end{equation}

\noindent
where $\rho$ is the energy density measured by a comoving observer
with the fluid, and $p_1$ and $p_2$ are the radial and tangential
pressures, respectively;
$u_\mu$ is the timelike four-velocity, and $x_\mu$ is the spacelike
unit vector orthogonal to $u_\mu$ and the angular directions, and
$g_{\mu\nu}$ is the metric of the background spacetime. 
We write

\begin{equation}
  u_\mu u^\mu = -1\, , \quad
  x_\mu x^\mu = 1\, , \quad \textrm{and} \quad
  u_\mu x^\mu = 0\, .
\end{equation}

The non-zero components of the Einstein equation for the
quantum-corrected spacetime lead to  

\begin{equation}
  T_0^0 = -\rho\, , \quad
  T_1^1 = p_1\, , \quad \textrm{and} \quad
  T_2^2 = T_3^3 = p_2\, .
\end{equation}

At this point the tetrad formalism (not Newman-Penrose formalism) is
typically used~\cite{Chen:2019iuo}. 
Perturbing the energy-momentum tensor in the tetrad frame and using
the constrains on $u^\mu$ and $x^\mu$ above leads to the vanishing
axial components 

\begin{equation}
\delta T_{(1)(0)} = \delta T_{(1)(2)} = \delta T_{(1)(3)} = 0
\end{equation}

\noindent
if the perturbed energy-momentum tensor is defined by an anisotropic  
fluid~\cite{Chen:2019iuo}. 
The round bracketed subscripts are tetrad-frame indices. 

Since the axial components of the perturbed energy-momentum tensor
vanish, the perturbation equation for the axial perturbations can be
derived from $R_{(a)(b)}|_\mathrm{axial} = 0$.
The vanishing of the axial components in the tetrad frame is not
equivalent to the vanishing of those of the Ricci tensor $R_{\mu\nu}$
on the coordinate basis.
In fact, the master equation of the axial perturbations derived from
the tetrad $R_{(a)(b)}$ is equivalent to that derived from the
linearized Einstein equation coupled to the energy-momentum tensor of
an anisotropic fluid. 

The vanishing of the axial components of the perturbed Ricci tensor
can be used to obtain two perturbation equations.
The perturbation equations can be solved to obtain a
Schr{\"o}dinger-like equation for the axial-gravitational
perturbations and an effective
potential~\cite{Bouhmadi-Lopez:2020oia,Yang:2023gas}.
We do not copy the derivation here.

On general grounds, the axial components of the perturbed
energy-momentum tensor of the anisotropic fluid vanish and are not
dependent on the metric functions.
The potential thus obtained in~\cite{Yang:2023gas} is applicable for
axial gravitational perturbations of any spherically symmetric static
spacetime.
In addition, it can be shown that the resulting effective potential is
equal to that used in this paper and given by~\cite{Moulin:2019ekf},
and elsewhere.

\section{Boundary conditions and an asymptotic solution\label{app:B}}

In this appendix we discuss the QNM boundary conditions and write down
an asymptotic solution.
The aim is to solve the \SH-like wave equation~(\ref{eq:schrodinger})
for the complex eigenfrequencies $\omega$. 
The potential is zero at the horizon and at spatial infinity, and the
\SH-like equation becomes a harmonic oscillator problem.
The boundary condition at the horizon is a wave that purely goes into
the black hole; representing classically that nothing comes out.
The other boundary condition at spatial infinity is a wave that purely
goes out; as nothing comes from outside the spacetime.
The boundary condition at spatial infinity is given as $r\to \infty$, 
$r_* \to r + (2m+r_0/2)\ln r$, which gives  
  
\begin{equation}
  \psi^{r\to\infty} \to e^{i\omega r_*} \to r^{i\omega(2m+r_0/2) }
  e^{i\omega r}\, .
\end{equation}

\noindent
Normalized to unity at the horizon, we write

\begin{equation}
  \psi^{r\to\infty}_{r\to r_h} \to r^{i\omega(2m+r_0/2)}
  e^{i\omega(r-2m)}\, .
\end{equation}

\noindent
The boundary condition at the event horizon is given as $r\to 2m$
($r_0\ne 2m$),

\begin{equation}
r_* \to  \frac{2m}{\sqrt{1 -\frac{r_0}{2m}}}\ln(r-2m)\, ,
\end{equation}

\noindent
which gives

\begin{equation}
  \psi^{r\to r_h} \to e^{-i\omega r_*} \to (r-2m)^\zeta\, ,
\end{equation}

\noindent
where

\begin{equation}
\zeta = \frac{-i\omega (2m)^{3/2}}{\sqrt{2m-r_0}}\, .
\end{equation}

\noindent
Normalized to unity at spatial infinity, we write

\begin{equation}
\psi^{r\to r_h}_{r\to \infty} \to e^{-i\omega r_*} \to (r-2m)^\zeta
r^{-\zeta}\, .
\end{equation}

\noindent
Combining the two asymptotic solutions, we form the asymptotic wave
function satisfying both boundary conditions

\begin{equation}
\psi(r)= r^{i\omega(2m+r_0/2) -\zeta} (r-2m)^\zeta e^{i\omega(r-2m)}\, .
\end{equation}

\noindent
The boundary conditions do not depend explicitly on $s$ or $\ell$,
so the above solution can be used for all spin and angular momentum
perturbations. 

\section{Tables of quasinormal mode frequencies\label{app:C}}

The following tables show some QNM frequencies for all integer spins
for $\ell=s$ and $\ell=s+1$, and spin-1/2 for $\ell=0$ and $\ell=1$. 
The fundamental and first few overtones are shown for the cases of
$r_0=0$ (Schwarzschild), and $r_0=0.3r_h$  and $r_0=0.5r_h$.
The WKB method has been used in the first two rows of each table
($n=0,1$).
Also shown in the middle section of each table, except for the cases
of $s=1/2$, are the fundamental and first five overtones using the PSM. 
The end section of the tables for integer spin (axial-gravitational),
show the fundamental and first five overtones using the CFM.

\begin{table}[htb]
\caption{Quasinormal mode frequencies $\omega$ for $s = 0$ and $\ell
= 0$ perturbations on the holonomy corrected black hole background 
metric. A mass $m=1/2$ has been used.}
\begin{ruledtabular}
\begin{tabular}{@{}cccc}
$n$ & $\omega(r_0=0)$ & $\omega(r_0=0.3r_h)$ &
$\omega(r_0=0.5r_h)$\\
\colrule
\multicolumn{4}{c}{WKB method}\\ 
0 & $0.221398 - 0.209668i$ & $0.210940 - 0.181256i$ & $0.201627 - 0.165232i$\\
1 & $0.173682 - 0.695890i$ & $0.172882 - 0.608171i$ & $0.150244 - 0.534780i$\\
\colrule
\multicolumn{4}{c}{pseudo-spectral method}\\
0 & $0.220910 - 0.209791i$ & $0.209951 - 0.182897i$ & $0.200799 - 0.164527i$\\
1 & $0.172234 - 0.696105i$ & $0.171098 - 0.600071i$ & $0.160438 - 0.534669i$\\
2 & $0.151489 - 1.202170i$ & $0.151642 - 1.027013i$ & $0.126270 - 0.921447i$\\
3 & $0.140902 - 1.707490i$ & $0.148772 - 1.455777i$ & $0.075090 - 1.315370i$\\
4 & $0.134771 - 2.211125i$ & $0.140174 - 1.918213i$ & $0.057686 - 1.817240i$\\
5 & $0.128360 - 2.715460i$ & $0.137054 - 2.307825i$ & $0.100769 - 2.211980i$\\
\colrule
\multicolumn{4}{c}{continued fraction method}\\
0 & $0.220910 - 0.209791i$ & $0.209951 - 0.182897i$ & $0.200799 - 0.164527i$\\
1 & $0.172238 - 0.696105i$ & $0.170562 - 0.599662i$ & $0.160438 - 0.534669i$\\
2 & $0.151484 - 1.202157i$ & $0.151873 - 1.032994i$ & $0.126272 - 0.921437i$\\
3 & $0.140820 - 1.707355i$ & $0.140359 - 1.465705i$ & $0.076061 - 1.315753i$\\
4 & $0.134149 - 2.211264i$ & $0.130799 - 1.897413i$ & $0.053619 - 1.815275i$\\
5 & $0.129483 - 2.714279i$ & $0.121348 - 2.328582i$ & $0.101169 - 2.213829i$\\
\end{tabular}
\end{ruledtabular}
\end{table}

\begin{table}[htb]
\caption{Quasinormal mode frequencies $\omega$ for $s = 0$ and $\ell
= 1$ perturbations on the holonomy corrected black hole background 
metric. A mass $m=1/2$ has been used.}
\begin{ruledtabular}
\begin{tabular}{@{}cccc}
$n$ & $\omega(r_0=0)$ & $\omega(r_0=0.3r_h)$ & $\omega(r_0=0.5r_h)$\\
\colrule
\multicolumn{4}{c}{WKB method}\\
0 & $0.585872 - 0.195320i$ & $0.582409 - 0.173800i$ & $0.579654 - 0.158246i$\\
1 & $0.528930 - 0.612519i$ & $0.541781 - 0.539669i$ & $0.547228 - 0.487774i$\\
\colrule
\multicolumn{4}{c}{pseudo-spectral method}\\
0 & $0.585872 - 0.195320i$ & $0.582405 - 0.173811i$ & $0.579649 - 0.158265i$\\
1 & $0.528897 - 0.612515i$ & $0.541621 - 0.539546i$ & $0.547089 - 0.487735i$\\
2 & $0.459079 - 1.080270i$ & $0.489270 - 0.940615i$ & $0.502191 - 0.843179i$\\
3 & $0.406517 - 1.576600i$ & $0.447087 - 1.364083i$ & $0.461392 - 1.216780i$\\
4 & $0.370218 - 2.081520i$ & $0.417983 - 1.793445i$ & $0.426854 - 1.597880i$\\
5 & $0.344152 - 2.588240i$ & $0.391861 - 2.234907i$ & $0.395550 - 1.981460i$\\
\colrule
\multicolumn{4}{c}{continued fraction method}\\
0 & $0.585872 - 0.195320i$ & $0.582405 - 0.173811i$ & $0.579649 - 0.158265i$\\
1 & $0.528897 - 0.612515i$ & $0.541621 - 0.539546i$ & $0.547089 - 0.487735i$\\
2 & $0.459079 - 1.080267i$ & $0.489294 - 0.940605i$ & $0.502191 - 0.843179i$\\
3 & $0.406517 - 1.576596i$ & $0.447379 - 1.364000i$ & $0.461392 - 1.216781i$\\
4 & $0.370218 - 2.081524i$ & $0.417232 - 1.795239i$ & $0.426854 - 1.597881i$\\
5 & $0.344154 - 2.588239i$ & $0.394909 - 2.228387i$ & $0.395550 - 1.981462i$\\
\end{tabular}
\end{ruledtabular}
\end{table}

\begin{table}[htb]
\caption{Quasinormal mode frequencies $\omega$ for $s = 1$ and $\ell
= 1$ perturbations on the holonomy corrected black hole background 
metric. A mass $m=1/2$ has been used.}
\begin{ruledtabular}
\begin{tabular}{@{}cccc}
$n$ & $\omega(r_0=0)$ & $\omega(r_0=0.3r_h)$ & $\omega(r_0=0.5r_h)$\\
\colrule
\multicolumn{4}{c}{WKB method}\\
0 & $0.496476 - 0.184951i$ & $0.506877 - 0.166933i$ & $0.513372 -0.152887i$\\
1 & $0.429027 - 0.587359i$ & $0.458977 - 0.523316i$ & $0.476678 -0.475129i$\\
\colrule
\multicolumn{4}{c}{pseudo-spectral method}\\
0 & $0.496527 - 0.184975i$ & $0.506890 - 0.166895i$ & $0.513377 -0.152855i$\\
1 & $0.429031 - 0.587335i$ & $0.459109 - 0.522645i$ & $0.476434 -0.474099i$\\
2 & $0.349547 - 1.050380i$ & $0.399830 - 0.920377i$ & $0.427459 -0.825862i$\\
3 & $0.292353 - 1.543820i$ & $0.353765 - 1.341634i$ & $0.385422 -1.197330i$\\
4 & $0.253105 - 2.045010i$ & $0.326140 - 1.771052i$ & $0.351646 -1.576040i$\\
5 & $0.224547 - 2.547800i$ & $0.283109 - 2.203791i$ & $0.322641
-1.956800i$\\
\colrule
\multicolumn{4}{c}{continued fraction method}\\
0 & $0.496527 - 0.184975i$ & $0.505594 - 0.168286i$ & $0.509494 -0.1564254i$\\
1 & $0.429031 - 0.587335i$ & $0.458776 - 0.527119i$ & $0.474312 -0.4851869i$\\
2 & $0.349547 - 1.050375i$ & $0.402147 - 0.927330i$ & $0.430014 -0.8430349i$\\
3 & $0.292353 - 1.543818i$ & $0.359743 - 1.350424i$ & $0.393691 -1.2176573i$\\
4 & $0.253108 - 2.045101i$ & $0.330506 - 1.780365i$ & $0.365655 -1.5975362i$\\
5 & $0.224506 - 2.547851i$ & $0.309577 - 2.211737i$ & $0.343004 -1.9781965i$\\
\end{tabular}
\end{ruledtabular}
\end{table}

\begin{table}[htb]
\caption{Quasinormal mode frequencies $\omega$ for $s = 1$ and $\ell
= 2$ perturbations on the holonomy corrected black hole background 
metric. A mass $m=1/2$ has been used.}
\begin{ruledtabular}
\begin{tabular}{@{}cccc}
$n$ & $\omega(r_0=0)$ & $\omega(r_0=0.3r_h)$ & $\omega(r_0=0.5r_h)$\\
\colrule
\multicolumn{4}{c}{WKB method}\\
0 & $0.915191 - 0.190009i$ & $0.921070 - 0.170350i$ & $0.924717 - 0.155639i$\\
1 & $0.873085 - 0.581420i$ & $0.891515 - 0.518850i$ & $0.901953 -
0.472429i$\\
\colrule
\multicolumn{4}{c}{pseudo-spectral method}\\
0 & $0.915191 - 0.190009i$ & $ 0.921070 - 0.170350i$ & $0.924716 - 0.155637i$\\
1 & $0.873085 - 0.581420i$ & $ 0.891517 - 0.518808i$ & $0.901934 - 0.472399i$\\
2 & $0.802373 - 1.003170i$ & $ 0.841413 - 0.887662i$ & $0.862549 - 0.803573i$\\
3 & $0.725190 - 1.460490i$ & $ 0.784976 - 1.280840i$ & $0.816205 - 1.152340i$\\
4 & $0.657473 - 1.943220i$ & $ 0.733214 - 1.693063i$ & $0.770903 - 1.515800i$\\
5 & $0.602986 - 2.439430i$ & $ 0.690796 - 2.116319i$ & $0.730157 - 1.888690i$\\
\colrule
\multicolumn{4}{c}{continued fraction method}\\
0 & $0.915191 - 0.190009i$ & $0.920497 - 0.170816i$ & $0.922872 - 0.156839i$\\
1 & $0.873085 - 0.581420i$ & $0.891121 - 0.520253i$ & $0.900361 - 0.476076i$\\
2 & $0.802373 - 1.003175i$ & $0.841506 - 0.890132i$ & $0.861747 - 0.809763i$\\
3 & $0.725190 - 1.460397i$ & $0.785928 - 1.284202i$ & $0.816798 - 1.160768i$\\
4 & $0.657473 - 1.943219i$ & $0.735330 - 1.697013i$ & $0.773338 - 1.525880i$\\
5 & $0.602986 - 2.439430i$ & $0.693438 - 2.120696i$ & $0.734635 - 1.899820i$\\
\end{tabular}
\end{ruledtabular}
\end{table}

\begin{table}[htb]
\caption{Quasinormal mode frequencies $\omega$ for $s = 2$ (axial) and $\ell
= 2$ perturbations on the holonomy corrected black hole background 
metric. A mass $m=1/2$ has been used.}
\begin{ruledtabular}
\begin{tabular}{@{}cccc}
$n$ & $\omega(r_0=0)$ & $\omega(r_0=0.3r_h)$ & $\omega(r_0=0.5r_h)$\\
\colrule
\multicolumn{4}{c}{WKB method}\\
0 & $0.747218 - 0.177952i$ & $0.748072 - 0.158889i$ & $0.748421 - 0.144849i$\\
1 & $0.693168 - 0.547883i$ & $0.711182 - 0.486901i$ & $0.720573 -
0.442139i$\\
\colrule
\multicolumn{4}{c}{pseudo-spectral method}\\
0 & $0.747343 - 0.177925i$ & $0.748069 - 0.158884i$ & $0.748438 - 0.144861i$\\
1 & $0.693422 - 0.547830i$ & $0.711257 - 0.486748i$ & $0.720861 - 0.442027i$\\
2 & $0.602107 - 0.956554i$ & $0.649620 - 0.841618i$ & $0.674269 - 0.758862i$\\
3 & $0.503010 - 1.410300i$ & $0.583014 - 1.227989i$ & $0.622339 - 1.099070i$\\
4 & $0.415029 - 1.893690i$ & $0.525807 - 1.636365i$ & $0.575360 - 1.457200i$\\
5 & $0.338598 - 2.391220i$ & $0.479007 - 2.059199i$ & $0.536349 - 1.825680i$\\
\colrule
\multicolumn{4}{c}{continued fraction method}\\
0 & $0.747343 - 0.177925i$ & $0.744542 - 0.160930i$ & $0.748034 - 0.150324i$\\
1 & $0.693422 - 0.547830i$ & $0.707555 - 0.493203i$ & $0.709728 - 0.458873i$\\
2 & $0.602107 - 0.956554i$ & $0.646457 - 0.852873i$ & $0.663411 - 0.787356i$\\
3 & $0.503010 - 1.410296i$ & $0.581673 - 1.243425i$ & $0.613906 - 1.137417i$\\
4 & $0.415029 - 1.893690i$ & $0.526739 - 1.655046i$ & $0.570861 - 1.502086i$\\
5 & $0.338599 - 2.391216i$ & $0.484189 - 2.077092i$ & $0.536773 - 1.874033i$\\
\end{tabular}
\end{ruledtabular}
\end{table}

\begin{table}[htb]
\caption{Quasinormal mode frequencies $\omega$ for $s = 2$ (polar) and $\ell
= 2$ perturbations on the holonomy corrected black hole background 
metric. A mass $m=1/2$ has been used.}
\begin{ruledtabular}
\begin{tabular}{@{}cccc}
$n$ & $\omega(r_0=0)$ & $\omega(r_0=0.3r_h)$ & $\omega(r_0=0.5r_h)$\\
\colrule
\multicolumn{4}{c}{WKB method}\\
0 & $0.747343 - 0.177925i$ & $0.751337 - 0.161732i$ & $0.754175 - 0.148937i$\\
1 & $0.693355 - 0.547887i$ & $0.711364 - 0.495572i$ & $0.722653 - 0.454531i$\\
\colrule
\multicolumn{4}{c}{pseudo-spectral method}\\
0 & $0.747343 - 0.177925i$ & $0.751336 -0.161733i$ & $0.754174 - 0.148939i$\\
1 & $0.693422 - 0.547830i$ & $0.711339 -0.495544i$ & $0.722707 - 0.454532i$\\
2 & $0.602107 - 0.956554i$ & $0.642989 -0.857438i$ & $0.668234 - 0.780900i$\\
3 & $0.503010 - 1.410300i$ & $0.565997 -1.253450i$ & $0.604769 - 1.133160i$\\
4 & $0.415029 - 1.893690i$ & $0.495057 -1.676309i$ & $0.544089 - 1.507670i$\\
5 & $0.338598 - 2.391220i$ & $0.436764 -2.120148i$ & $0.492696 - 1.897920i$\\
\end{tabular}
\end{ruledtabular}
\end{table}

\begin{table}[htb]
\caption{Quasinormal mode frequencies $\omega$ for $s = 2$ (axial) and $\ell
= 3$ perturbations on the holonomy corrected black hole background 
metric. A mass $m=1/2$ has been used.}
\begin{ruledtabular}
\begin{tabular}{@{}cccc}
$n$ & $\omega(r_0=0)$ & $\omega(r_0=0.3r_h)$ & $\omega(r_0=0.5r_h)$\\
\colrule
\multicolumn{4}{c}{WKB method}\\
0 & $1.198890 - 0.185406i$ & $1.200340 - 0.166139i$ & $1.201200 - 0.151817i$\\
1 & $1.165290 - 0.562597i$ & $1.176720 - 0.502876i$ & $1.183060 -
0.458652i$\\
\colrule
\multicolumn{4}{c}{pseudo-spectral method}\\
0 & $1.198890 - 0.185406i$ & $1.201233 - 0.166756i$ & $1.201200 - 0.151817i$\\
1 & $1.165290 - 0.562596i$ & $1.177157 - 0.504736i$ & $1.183040 - 0.458653i$\\
2 & $1.103370 - 0.958186i$ & $1.132747 - 0.855406i$ & $1.149390 - 0.774523i$\\
3 & $1.023920 - 1.380670i$ & $1.075161 - 1.224438i$ & $1.105170 - 1.103430i$\\
4 & $0.940348 - 1.831300i$ & $1.013056 - 1.613060i$ & $1.056270 - 1.446480i$\\
5 & $0.862773 - 2.304300i$ & $0.953378 - 2.018204i$ & $1.007660 - 1.802000i$\\
\colrule
\multicolumn{4}{c}{continued fraction method}\\
0 & $1.198887 - 0.185406i$ & $1.198708 - 0.167057i$ & $1.196096 - 0.154212i$\\
1 & $1.165288 - 0.562596i$ & $1.175183 - 0.505692i$ & $1.177996 - 0.465935i$\\
2 & $1.103370 - 0.958186i$ & $1.132016 - 0.857111i$ & $1.144706 - 0.786875i$\\
3 & $1.023924 - 1.380674i$ & $1.076545 - 1.226852i$ & $1.101467 - 1.120852i$\\
4 & $0.940348 - 1.831299i$ & $1.017461 - 1.615873i$ & $1.054309 - 1.468575i$\\
5 & $0.862773 - 2.304303i$ & $0.961598 - 2.020917i$ & $1.008081 - 1.827995i$\\
\end{tabular}
\end{ruledtabular}
\end{table}

\begin{table}[htb]
\caption{Quasinormal mode frequencies $\omega$ for $s = 2$ (polar) and $\ell
= 3$ perturbations on the holonomy corrected black hole background 
 metric. A mass $m=1/2$ has been used.}
\begin{ruledtabular}
\begin{tabular}{@{}cccc}
$n$ & $\omega(r_0=0)$ & $\omega(r_0=0.3r_h)$ & $\omega(r_0=0.5r_h)$\\
\colrule
\multicolumn{4}{c}{WKB method}\\
0 & $1.198887 - 0.185406i$ & $1.201230 - 0.166756i$ & $1.20274 - 0.152685i$\\
1 & $1.165290 - 0.562596i$ & $1.177160 - 0.504743i$ & $1.18403 - 0.461288i$\\
\colrule
\multicolumn{4}{c}{pseudo-spectral method}\\
0 & $1.198890 - 0.185406i$ & $1.201233 - 0.166756i$ & $1.20274 - 0.152685i$\\
1 & $1.165290 - 0.562596i$ & $1.177157 - 0.504736i$ & $1.18403 - 0.461276i$\\
2 & $1.103370 - 0.958186i$ & $1.132747 - 0.855406i$ & $1.14928 - 0.778963i$\\
3 & $1.023920 - 1.380670i$ & $1.075161 - 1.224438i$ & $1.10341 - 1.109800i$\\
4 & $0.940348 - 1.831300i$ & $1.013056 - 1.613060i$ & $1.05234 - 1.454960i$\\
5 & $0.862773 - 2.304300i$ & $0.953378 - 2.018204i$ & $1.00110 - 1.812800i$\\
\end{tabular}
\end{ruledtabular}
\end{table}

\begin{table}[htb]
\caption{Quasinormal mode frequencies $\omega$ for $s = 1/2$ and $\ell
= 0$ perturbations on the holonomy corrected black hole background 
metric. A mass $m=1/2$ has been used.}
\begin{ruledtabular}
\begin{tabular}{@{}cccc}
$n$ & $\omega(r_0=0)$ & $\omega(r_0=0.3r_h)$ &$\omega(r_0=0.5r_h)$\\
\colrule
\multicolumn{4}{c}{WKB method}\\
0 & $0.366015 - 0.193958i$ & $0.370094 - 0.172478i$ & $0.370772 - 0.156146i$\\
1 & $0.296029 - 0.634352i$ & $0.310615 - 0.551149i$ & $0.328316 - 0.491751i$\\
\end{tabular}
\end{ruledtabular}
\end{table}

\begin{table}[htb]
\caption{Quasinormal mode frequencies $\omega$ for $s =1/ 2$ and $\ell
= 1$ perturbations on the holonomy corrected black hole background 
metric. A mass $m=1/2$ has been used.}
\begin{ruledtabular}
\begin{tabular}{@{}cccc}
$n$ & $\omega(r_0=0)$ & $\omega(r_0=0.3r_h)$ & $\omega(r_0=0.5r_h)$\\
\colrule
\multicolumn{4}{c}{WKB method}\\
0 & $0.760074 - 0.192810i$ & $0.762146 - 0.172211i$ & $0.763653 - 0.157223i$\\
1 & $0.711660 - 0.594999i$ & $0.726414 - 0.528276i$ & $0.737744 - 0.478374i$\\
\end{tabular}
\end{ruledtabular}
\end{table}

\clearpage
\bibliography{gingrich}

\begin{thebibliography}{65}%
\makeatletter
\providecommand \@ifxundefined [1]{%
 \@ifx{#1\undefined}
}%
\providecommand \@ifnum [1]{%
 \ifnum #1\expandafter \@firstoftwo
 \else \expandafter \@secondoftwo
 \fi
}%
\providecommand \@ifx [1]{%
 \ifx #1\expandafter \@firstoftwo
 \else \expandafter \@secondoftwo
 \fi
}%
\providecommand \natexlab [1]{#1}%
\providecommand \enquote  [1]{``#1''}%
\providecommand \bibnamefont  [1]{#1}%
\providecommand \bibfnamefont [1]{#1}%
\providecommand \citenamefont [1]{#1}%
\providecommand \href@noop [0]{\@secondoftwo}%
\providecommand \href [0]{\begingroup \@sanitize@url \@href}%
\providecommand \@href[1]{\@@startlink{#1}\@@href}%
\providecommand \@@href[1]{\endgroup#1\@@endlink}%
\providecommand \@sanitize@url [0]{\catcode `\\12\catcode `\$12\catcode
  `\&12\catcode `\#12\catcode `\^12\catcode `\_12\catcode `\%12\relax}%
\providecommand \@@startlink[1]{}%
\providecommand \@@endlink[0]{}%
\providecommand \url  [0]{\begingroup\@sanitize@url \@url }%
\providecommand \@url [1]{\endgroup\@href {#1}{\urlprefix }}%
\providecommand \urlprefix  [0]{URL }%
\providecommand \Eprint [0]{\href }%
\providecommand \doibase [0]{https://doi.org/}%
\providecommand \selectlanguage [0]{\@gobble}%
\providecommand \bibinfo  [0]{\@secondoftwo}%
\providecommand \bibfield  [0]{\@secondoftwo}%
\providecommand \translation [1]{[#1]}%
\providecommand \BibitemOpen [0]{}%
\providecommand \bibitemStop [0]{}%
\providecommand \bibitemNoStop [0]{.\EOS\space}%
\providecommand \EOS [0]{\spacefactor3000\relax}%
\providecommand \BibitemShut  [1]{\csname bibitem#1\endcsname}%
\let\auto@bib@innerbib\@empty
\bibitem [{\citenamefont {Perez}(2017)}]{Perez:2017cmj}%
  \BibitemOpen
  \bibfield  {author} {\bibinfo {author} {\bibfnamefont {A.}~\bibnamefont
  {Perez}},\ }\href {https://doi.org/10.1088/1361-6633/aa7e14} {\bibfield
  {journal} {\bibinfo  {journal} {Rept. Prog. Phys.}\ }\textbf {\bibinfo
  {volume} {80}},\ \bibinfo {pages} {126901} (\bibinfo {year} {2017})},\
  \Eprint {https://arxiv.org/abs/1703.09149} {arXiv:1703.09149 [gr-qc]}
  \BibitemShut {NoStop}%
\bibitem [{\citenamefont {Zhang}(2023)}]{Zhang:2023yps}%
  \BibitemOpen
  \bibfield  {author} {\bibinfo {author} {\bibfnamefont {X.}~\bibnamefont
  {Zhang}},\ }\href {https://doi.org/10.3390/universe9070313} {\bibfield
  {journal} {\bibinfo  {journal} {Universe}\ }\textbf {\bibinfo {volume} {9}},\
  \bibinfo {pages} {313} (\bibinfo {year} {2023})},\ \Eprint
  {https://arxiv.org/abs/2308.10184} {arXiv:2308.10184 [gr-qc]} \BibitemShut
  {NoStop}%
\bibitem [{\citenamefont {Modesto}(2010)}]{Modesto:2008im}%
  \BibitemOpen
  \bibfield  {author} {\bibinfo {author} {\bibfnamefont {L.}~\bibnamefont
  {Modesto}},\ }\href {https://doi.org/10.1007/s10773-010-0346-x} {\bibfield
  {journal} {\bibinfo  {journal} {Int. J. Theor. Phys.}\ }\textbf {\bibinfo
  {volume} {49}},\ \bibinfo {pages} {1649} (\bibinfo {year} {2010})},\ \Eprint
  {https://arxiv.org/abs/0811.2196} {arXiv:0811.2196 [gr-qc]} \BibitemShut
  {NoStop}%
\bibitem [{\citenamefont {Peltola}\ and\ \citenamefont
  {Kunstatter}(2009)}]{Peltola:2008pa}%
  \BibitemOpen
  \bibfield  {author} {\bibinfo {author} {\bibfnamefont {A.}~\bibnamefont
  {Peltola}}\ and\ \bibinfo {author} {\bibfnamefont {G.}~\bibnamefont
  {Kunstatter}},\ }\href {https://doi.org/10.1103/PhysRevD.79.061501}
  {\bibfield  {journal} {\bibinfo  {journal} {Phys. Rev. D}\ }\textbf {\bibinfo
  {volume} {79}},\ \bibinfo {pages} {061501} (\bibinfo {year} {2009})},\
  \Eprint {https://arxiv.org/abs/0811.3240} {arXiv:0811.3240 [gr-qc]}
  \BibitemShut {NoStop}%
\bibitem [{\citenamefont {Bodendorfer}\ \emph {et~al.}(2019)\citenamefont
  {Bodendorfer}, \citenamefont {Mele},\ and\ \citenamefont
  {M\"unch}}]{Bodendorfer:2019cyv}%
  \BibitemOpen
  \bibfield  {author} {\bibinfo {author} {\bibfnamefont {N.}~\bibnamefont
  {Bodendorfer}}, \bibinfo {author} {\bibfnamefont {F.~M.}\ \bibnamefont
  {Mele}},\ and\ \bibinfo {author} {\bibfnamefont {J.}~\bibnamefont
  {M\"unch}},\ }\href {https://doi.org/10.1088/1361-6382/ab3f16} {\bibfield
  {journal} {\bibinfo  {journal} {Class. Quant. Grav.}\ }\textbf {\bibinfo
  {volume} {36}},\ \bibinfo {pages} {195015} (\bibinfo {year} {2019})},\
  \Eprint {https://arxiv.org/abs/1902.04542} {arXiv:1902.04542 [gr-qc]}
  \BibitemShut {NoStop}%
\bibitem [{\citenamefont {Bodendorfer}\ \emph
  {et~al.}(2021{\natexlab{a}})\citenamefont {Bodendorfer}, \citenamefont
  {Mele},\ and\ \citenamefont {M\"unch}}]{Bodendorfer:2019nvy}%
  \BibitemOpen
  \bibfield  {author} {\bibinfo {author} {\bibfnamefont {N.}~\bibnamefont
  {Bodendorfer}}, \bibinfo {author} {\bibfnamefont {F.~M.}\ \bibnamefont
  {Mele}},\ and\ \bibinfo {author} {\bibfnamefont {J.}~\bibnamefont
  {M\"unch}},\ }\href {https://doi.org/10.1016/j.physletb.2021.136390}
  {\bibfield  {journal} {\bibinfo  {journal} {Phys. Lett. B}\ }\textbf
  {\bibinfo {volume} {819}},\ \bibinfo {pages} {136390} (\bibinfo {year}
  {2021}{\natexlab{a}})},\ \Eprint {https://arxiv.org/abs/1911.12646}
  {arXiv:1911.12646 [gr-qc]} \BibitemShut {NoStop}%
\bibitem [{\citenamefont {Bodendorfer}\ \emph
  {et~al.}(2021{\natexlab{b}})\citenamefont {Bodendorfer}, \citenamefont
  {Mele},\ and\ \citenamefont {M\"unch}}]{Bodendorfer:2019jay}%
  \BibitemOpen
  \bibfield  {author} {\bibinfo {author} {\bibfnamefont {N.}~\bibnamefont
  {Bodendorfer}}, \bibinfo {author} {\bibfnamefont {F.~M.}\ \bibnamefont
  {Mele}},\ and\ \bibinfo {author} {\bibfnamefont {J.}~\bibnamefont
  {M\"unch}},\ }\href {https://doi.org/10.1088/1361-6382/abe05d} {\bibfield
  {journal} {\bibinfo  {journal} {Class. Quant. Grav.}\ }\textbf {\bibinfo
  {volume} {38}},\ \bibinfo {pages} {095002} (\bibinfo {year}
  {2021}{\natexlab{b}})},\ \Eprint {https://arxiv.org/abs/1912.00774}
  {arXiv:1912.00774 [gr-qc]} \BibitemShut {NoStop}%
\bibitem [{\citenamefont {Ashtekar}\ \emph
  {et~al.}(2018{\natexlab{a}})\citenamefont {Ashtekar}, \citenamefont
  {Olmedo},\ and\ \citenamefont {Singh}}]{Ashtekar:2018lag}%
  \BibitemOpen
  \bibfield  {author} {\bibinfo {author} {\bibfnamefont {A.}~\bibnamefont
  {Ashtekar}}, \bibinfo {author} {\bibfnamefont {J.}~\bibnamefont {Olmedo}},\
  and\ \bibinfo {author} {\bibfnamefont {P.}~\bibnamefont {Singh}},\ }\href
  {https://doi.org/10.1103/PhysRevLett.121.241301} {\bibfield  {journal}
  {\bibinfo  {journal} {Phys. Rev. Lett.}\ }\textbf {\bibinfo {volume} {121}},\
  \bibinfo {pages} {241301} (\bibinfo {year} {2018}{\natexlab{a}})},\ \Eprint
  {https://arxiv.org/abs/1806.00648} {arXiv:1806.00648 [gr-qc]} \BibitemShut
  {NoStop}%
\bibitem [{\citenamefont {Ashtekar}\ \emph
  {et~al.}(2018{\natexlab{b}})\citenamefont {Ashtekar}, \citenamefont
  {Olmedo},\ and\ \citenamefont {Singh}}]{Ashtekar:2018cay}%
  \BibitemOpen
  \bibfield  {author} {\bibinfo {author} {\bibfnamefont {A.}~\bibnamefont
  {Ashtekar}}, \bibinfo {author} {\bibfnamefont {J.}~\bibnamefont {Olmedo}},\
  and\ \bibinfo {author} {\bibfnamefont {P.}~\bibnamefont {Singh}},\ }\href
  {https://doi.org/10.1103/PhysRevD.98.126003} {\bibfield  {journal} {\bibinfo
  {journal} {Phys. Rev. D}\ }\textbf {\bibinfo {volume} {98}},\ \bibinfo
  {pages} {126003} (\bibinfo {year} {2018}{\natexlab{b}})},\ \Eprint
  {https://arxiv.org/abs/1806.02406} {arXiv:1806.02406 [gr-qc]} \BibitemShut
  {NoStop}%
\bibitem [{\citenamefont {Ashtekar}\ and\ \citenamefont
  {Olmedo}(2020)}]{Ashtekar:2020ckv}%
  \BibitemOpen
  \bibfield  {author} {\bibinfo {author} {\bibfnamefont {A.}~\bibnamefont
  {Ashtekar}}\ and\ \bibinfo {author} {\bibfnamefont {J.}~\bibnamefont
  {Olmedo}},\ }\href {https://doi.org/10.1142/S0218271820500765} {\bibfield
  {journal} {\bibinfo  {journal} {Int. J. Mod. Phys. D}\ }\textbf {\bibinfo
  {volume} {29}},\ \bibinfo {pages} {2050076} (\bibinfo {year} {2020})},\
  \Eprint {https://arxiv.org/abs/2005.02309} {arXiv:2005.02309 [gr-qc]}
  \BibitemShut {NoStop}%
\bibitem [{\citenamefont {Gambini}\ and\ \citenamefont
  {Pullin}(2013)}]{Gambini:2013ooa}%
  \BibitemOpen
  \bibfield  {author} {\bibinfo {author} {\bibfnamefont {R.}~\bibnamefont
  {Gambini}}\ and\ \bibinfo {author} {\bibfnamefont {J.}~\bibnamefont
  {Pullin}},\ }\href {https://doi.org/10.1103/PhysRevLett.110.211301}
  {\bibfield  {journal} {\bibinfo  {journal} {Phys. Rev. Lett.}\ }\textbf
  {\bibinfo {volume} {110}},\ \bibinfo {pages} {211301} (\bibinfo {year}
  {2013})},\ \Eprint {https://arxiv.org/abs/1302.5265} {arXiv:1302.5265
  [gr-qc]} \BibitemShut {NoStop}%
\bibitem [{\citenamefont {Gambini}\ \emph {et~al.}(2014)\citenamefont
  {Gambini}, \citenamefont {Olmedo},\ and\ \citenamefont
  {Pullin}}]{Gambini:2013hna}%
  \BibitemOpen
  \bibfield  {author} {\bibinfo {author} {\bibfnamefont {R.}~\bibnamefont
  {Gambini}}, \bibinfo {author} {\bibfnamefont {J.}~\bibnamefont {Olmedo}},\
  and\ \bibinfo {author} {\bibfnamefont {J.}~\bibnamefont {Pullin}},\ }\href
  {https://doi.org/10.1088/0264-9381/31/9/095009} {\bibfield  {journal}
  {\bibinfo  {journal} {Class. Quant. Grav.}\ }\textbf {\bibinfo {volume}
  {31}},\ \bibinfo {pages} {095009} (\bibinfo {year} {2014})},\ \Eprint
  {https://arxiv.org/abs/1310.5996} {arXiv:1310.5996 [gr-qc]} \BibitemShut
  {NoStop}%
\bibitem [{\citenamefont {Gambini}\ \emph {et~al.}(2020)\citenamefont
  {Gambini}, \citenamefont {Olmedo},\ and\ \citenamefont
  {Pullin}}]{Gambini:2020nsf}%
  \BibitemOpen
  \bibfield  {author} {\bibinfo {author} {\bibfnamefont {R.}~\bibnamefont
  {Gambini}}, \bibinfo {author} {\bibfnamefont {J.}~\bibnamefont {Olmedo}},\
  and\ \bibinfo {author} {\bibfnamefont {J.}~\bibnamefont {Pullin}},\ }\href
  {https://doi.org/10.1088/1361-6382/aba842} {\bibfield  {journal} {\bibinfo
  {journal} {Class. Quant. Grav.}\ }\textbf {\bibinfo {volume} {37}},\ \bibinfo
  {pages} {205012} (\bibinfo {year} {2020})},\ \Eprint
  {https://arxiv.org/abs/2006.01513} {arXiv:2006.01513 [gr-qc]} \BibitemShut
  {NoStop}%
\bibitem [{\citenamefont {Kelly}\ \emph {et~al.}(2020)\citenamefont {Kelly},
  \citenamefont {Santacruz},\ and\ \citenamefont
  {Wilson-Ewing}}]{Kelly:2020uwj}%
  \BibitemOpen
  \bibfield  {author} {\bibinfo {author} {\bibfnamefont {J.~G.}\ \bibnamefont
  {Kelly}}, \bibinfo {author} {\bibfnamefont {R.}~\bibnamefont {Santacruz}},\
  and\ \bibinfo {author} {\bibfnamefont {E.}~\bibnamefont {Wilson-Ewing}},\
  }\href {https://doi.org/10.1103/PhysRevD.102.106024} {\bibfield  {journal}
  {\bibinfo  {journal} {Phys. Rev. D}\ }\textbf {\bibinfo {volume} {102}},\
  \bibinfo {pages} {106024} (\bibinfo {year} {2020})},\ \Eprint
  {https://arxiv.org/abs/2006.09302} {arXiv:2006.09302 [gr-qc]} \BibitemShut
  {NoStop}%
\bibitem [{\citenamefont {Kelly}\ \emph {et~al.}(2021)\citenamefont {Kelly},
  \citenamefont {Santacruz},\ and\ \citenamefont
  {Wilson-Ewing}}]{Kelly:2020lec}%
  \BibitemOpen
  \bibfield  {author} {\bibinfo {author} {\bibfnamefont {J.~G.}\ \bibnamefont
  {Kelly}}, \bibinfo {author} {\bibfnamefont {R.}~\bibnamefont {Santacruz}},\
  and\ \bibinfo {author} {\bibfnamefont {E.}~\bibnamefont {Wilson-Ewing}},\
  }\href {https://doi.org/10.1088/1361-6382/abd3e2} {\bibfield  {journal}
  {\bibinfo  {journal} {Class. Quant. Grav.}\ }\textbf {\bibinfo {volume}
  {38}},\ \bibinfo {pages} {04LT01} (\bibinfo {year} {2021})},\ \Eprint
  {https://arxiv.org/abs/2006.09325} {arXiv:2006.09325 [gr-qc]} \BibitemShut
  {NoStop}%
\bibitem [{\citenamefont {Alonso-Bardaji}\ \emph
  {et~al.}(2022{\natexlab{a}})\citenamefont {Alonso-Bardaji}, \citenamefont
  {Brizuela},\ and\ \citenamefont {Vera}}]{Alonso-Bardaji:2021yls}%
  \BibitemOpen
  \bibfield  {author} {\bibinfo {author} {\bibfnamefont {A.}~\bibnamefont
  {Alonso-Bardaji}}, \bibinfo {author} {\bibfnamefont {D.}~\bibnamefont
  {Brizuela}},\ and\ \bibinfo {author} {\bibfnamefont {R.}~\bibnamefont
  {Vera}},\ }\href {https://doi.org/10.1016/j.physletb.2022.137075} {\bibfield
  {journal} {\bibinfo  {journal} {Phys. Lett. B}\ }\textbf {\bibinfo {volume}
  {829}},\ \bibinfo {pages} {137075} (\bibinfo {year} {2022}{\natexlab{a}})},\
  \Eprint {https://arxiv.org/abs/2112.12110} {arXiv:2112.12110 [gr-qc]}
  \BibitemShut {NoStop}%
\bibitem [{\citenamefont {Alonso-Bardaji}\ \emph
  {et~al.}(2022{\natexlab{b}})\citenamefont {Alonso-Bardaji}, \citenamefont
  {Brizuela},\ and\ \citenamefont {Vera}}]{Alonso-Bardaji:2022ear}%
  \BibitemOpen
  \bibfield  {author} {\bibinfo {author} {\bibfnamefont {A.}~\bibnamefont
  {Alonso-Bardaji}}, \bibinfo {author} {\bibfnamefont {D.}~\bibnamefont
  {Brizuela}},\ and\ \bibinfo {author} {\bibfnamefont {R.}~\bibnamefont
  {Vera}},\ }\href {https://doi.org/10.1103/PhysRevD.106.024035} {\bibfield
  {journal} {\bibinfo  {journal} {Phys. Rev. D}\ }\textbf {\bibinfo {volume}
  {106}},\ \bibinfo {pages} {024035} (\bibinfo {year} {2022}{\natexlab{b}})},\
  \Eprint {https://arxiv.org/abs/2205.02098} {arXiv:2205.02098 [gr-qc]}
  \BibitemShut {NoStop}%
\bibitem [{\citenamefont {Abbott}\ \emph {et~al.}(2016)\citenamefont {Abbott}
  \emph {et~al.}}]{LIGOScientific:2016aoc}%
  \BibitemOpen
  \bibfield  {author} {\bibinfo {author} {\bibfnamefont {B.~P.}\ \bibnamefont
  {Abbott}} \emph {et~al.} (\bibinfo {collaboration} {LIGO Scientific,
  Virgo}),\ }\href {https://doi.org/10.1103/PhysRevLett.116.061102} {\bibfield
  {journal} {\bibinfo  {journal} {Phys. Rev. Lett.}\ }\textbf {\bibinfo
  {volume} {116}},\ \bibinfo {pages} {061102} (\bibinfo {year} {2016})},\
  \Eprint {https://arxiv.org/abs/1602.03837} {arXiv:1602.03837 [gr-qc]}
  \BibitemShut {NoStop}%
\bibitem [{\citenamefont {Abbott}\ \emph {et~al.}(2022)\citenamefont {Abbott}
  \emph {et~al.}}]{KAGRA:2022twx}%
  \BibitemOpen
  \bibfield  {author} {\bibinfo {author} {\bibfnamefont {R.}~\bibnamefont
  {Abbott}} \emph {et~al.} (\bibinfo {collaboration} {KAGRA, VIRGO, LIGO
  Scientific}),\ }\href {https://doi.org/10.1093/ptep/ptac073} {\bibfield
  {journal} {\bibinfo  {journal} {PTEP}\ }\textbf {\bibinfo {volume} {2022}},\
  \bibinfo {pages} {063F01} (\bibinfo {year} {2022})},\ \Eprint
  {https://arxiv.org/abs/2203.01270} {arXiv:2203.01270 [gr-qc]} \BibitemShut
  {NoStop}%
\bibitem [{\citenamefont {Abbott}\ \emph {et~al.}(2023)\citenamefont {Abbott}
  \emph {et~al.}}]{KAGRA:2023pio}%
  \BibitemOpen
  \bibfield  {author} {\bibinfo {author} {\bibfnamefont {R.}~\bibnamefont
  {Abbott}} \emph {et~al.} (\bibinfo {collaboration} {KAGRA, VIRGO, LIGO
  Scientific}),\ }\href {https://doi.org/10.3847/1538-4365/acdc9f} {\bibfield
  {journal} {\bibinfo  {journal} {Astrophys. J. Suppl.}\ }\textbf {\bibinfo
  {volume} {267}},\ \bibinfo {pages} {29} (\bibinfo {year} {2023})},\ \Eprint
  {https://arxiv.org/abs/2302.03676} {arXiv:2302.03676 [gr-qc]} \BibitemShut
  {NoStop}%
\bibitem [{\citenamefont {Moreira}\ \emph {et~al.}(2023)\citenamefont
  {Moreira}, \citenamefont {Lima~Junior}, \citenamefont {Crispino},\ and\
  \citenamefont {Herdeiro}}]{Moreira:2023cxy}%
  \BibitemOpen
  \bibfield  {author} {\bibinfo {author} {\bibfnamefont {Z.~S.}\ \bibnamefont
  {Moreira}}, \bibinfo {author} {\bibfnamefont {H.~C.~D.}\ \bibnamefont
  {Lima~Junior}}, \bibinfo {author} {\bibfnamefont {L.~C.~B.}\ \bibnamefont
  {Crispino}},\ and\ \bibinfo {author} {\bibfnamefont {C.~A.~R.}\ \bibnamefont
  {Herdeiro}},\ }\href {https://doi.org/10.1103/PhysRevD.107.104016} {\bibfield
   {journal} {\bibinfo  {journal} {Phys. Rev. D}\ }\textbf {\bibinfo {volume}
  {107}},\ \bibinfo {pages} {104016} (\bibinfo {year} {2023})},\ \Eprint
  {https://arxiv.org/abs/2302.14722} {arXiv:2302.14722 [gr-qc]} \BibitemShut
  {NoStop}%
\bibitem [{\citenamefont {Fu}\ \emph {et~al.}(2024)\citenamefont {Fu},
  \citenamefont {Zhang}, \citenamefont {Liu}, \citenamefont {Kuang},\ and\
  \citenamefont {Wu}}]{Fu:2023drp}%
  \BibitemOpen
  \bibfield  {author} {\bibinfo {author} {\bibfnamefont {G.}~\bibnamefont
  {Fu}}, \bibinfo {author} {\bibfnamefont {D.}~\bibnamefont {Zhang}}, \bibinfo
  {author} {\bibfnamefont {P.}~\bibnamefont {Liu}}, \bibinfo {author}
  {\bibfnamefont {X.-M.}\ \bibnamefont {Kuang}},\ and\ \bibinfo {author}
  {\bibfnamefont {J.-P.}\ \bibnamefont {Wu}},\ }\href
  {https://doi.org/10.1103/PhysRevD.109.026010} {\bibfield  {journal} {\bibinfo
   {journal} {Phys. Rev. D}\ }\textbf {\bibinfo {volume} {109}},\ \bibinfo
  {pages} {026010} (\bibinfo {year} {2024})},\ \Eprint
  {https://arxiv.org/abs/2301.08421} {arXiv:2301.08421 [gr-qc]} \BibitemShut
  {NoStop}%
\bibitem [{\citenamefont {Soares}\ \emph {et~al.}(2023)\citenamefont {Soares},
  \citenamefont {Pereira}, \citenamefont {Vit\'oria},\ and\ \citenamefont
  {Rocha}}]{Soares:2023uup}%
  \BibitemOpen
  \bibfield  {author} {\bibinfo {author} {\bibfnamefont {A.~R.}\ \bibnamefont
  {Soares}}, \bibinfo {author} {\bibfnamefont {C.~F.~S.}\ \bibnamefont
  {Pereira}}, \bibinfo {author} {\bibfnamefont {R.~L.~L.}\ \bibnamefont
  {Vit\'oria}},\ and\ \bibinfo {author} {\bibfnamefont {E.~M.}\ \bibnamefont
  {Rocha}},\ }\href {https://doi.org/10.1103/PhysRevD.108.124024} {\bibfield
  {journal} {\bibinfo  {journal} {Phys. Rev. D}\ }\textbf {\bibinfo {volume}
  {108}},\ \bibinfo {pages} {124024} (\bibinfo {year} {2023})},\ \Eprint
  {https://arxiv.org/abs/2309.05106} {arXiv:2309.05106 [gr-qc]} \BibitemShut
  {NoStop}%
\bibitem [{\citenamefont {Junior}\ \emph {et~al.}(2024)\citenamefont {Junior},
  \citenamefont {Lobo}, \citenamefont {Rodrigues},\ and\ \citenamefont
  {Vieira}}]{Junior:2023xgl}%
  \BibitemOpen
  \bibfield  {author} {\bibinfo {author} {\bibfnamefont {E.~L.~B.}\
  \bibnamefont {Junior}}, \bibinfo {author} {\bibfnamefont {F.~S.~N.}\
  \bibnamefont {Lobo}}, \bibinfo {author} {\bibfnamefont {M.~E.}\ \bibnamefont
  {Rodrigues}},\ and\ \bibinfo {author} {\bibfnamefont {H.~A.}\ \bibnamefont
  {Vieira}},\ }\href {https://doi.org/10.1103/PhysRevD.109.024004} {\bibfield
  {journal} {\bibinfo  {journal} {Phys. Rev. D}\ }\textbf {\bibinfo {volume}
  {109}},\ \bibinfo {pages} {024004} (\bibinfo {year} {2024})},\ \Eprint
  {https://arxiv.org/abs/2309.02658} {arXiv:2309.02658 [gr-qc]} \BibitemShut
  {NoStop}%
\bibitem [{\citenamefont {Chandrasekhar}(1983)}]{Chandrasekhar}%
  \BibitemOpen
  \bibfield  {author} {\bibinfo {author} {\bibfnamefont {S.}~\bibnamefont
  {Chandrasekhar}},\ }\href@noop {} {\emph {\bibinfo {title} {The mathematical
  theory of black holes}}}\ (\bibinfo  {publisher} {Oxford University Press},\
  \bibinfo {year} {1983})\BibitemShut {NoStop}%
\bibitem [{\citenamefont {Chandrasekhar}(1979)}]{chandrasekharRN}%
  \BibitemOpen
  \bibfield  {author} {\bibinfo {author} {\bibfnamefont {S.}~\bibnamefont
  {Chandrasekhar}},\ }\href {http://www.jstor.org/stable/79661} {\bibfield
  {journal} {\bibinfo  {journal} {Proceedings of the Royal Society of London.
  Series A, Mathematical and Physical Sciences}\ }\textbf {\bibinfo {volume}
  {365}},\ \bibinfo {pages} {453} (\bibinfo {year} {1979})}\BibitemShut
  {NoStop}%
\bibitem [{\citenamefont {Pani}\ \emph
  {et~al.}(2013{\natexlab{a}})\citenamefont {Pani}, \citenamefont {Berti},\
  and\ \citenamefont {Gualtieri}}]{Pani:2013ija}%
  \BibitemOpen
  \bibfield  {author} {\bibinfo {author} {\bibfnamefont {P.}~\bibnamefont
  {Pani}}, \bibinfo {author} {\bibfnamefont {E.}~\bibnamefont {Berti}},\ and\
  \bibinfo {author} {\bibfnamefont {L.}~\bibnamefont {Gualtieri}},\ }\href
  {https://doi.org/10.1103/PhysRevLett.110.241103} {\bibfield  {journal}
  {\bibinfo  {journal} {Phys. Rev. Lett.}\ }\textbf {\bibinfo {volume} {110}},\
  \bibinfo {pages} {241103} (\bibinfo {year} {2013}{\natexlab{a}})},\ \Eprint
  {https://arxiv.org/abs/1304.1160} {arXiv:1304.1160 [gr-qc]} \BibitemShut
  {NoStop}%
\bibitem [{\citenamefont {Pani}\ \emph
  {et~al.}(2013{\natexlab{b}})\citenamefont {Pani}, \citenamefont {Berti},\
  and\ \citenamefont {Gualtieri}}]{Pani:2013wsa}%
  \BibitemOpen
  \bibfield  {author} {\bibinfo {author} {\bibfnamefont {P.}~\bibnamefont
  {Pani}}, \bibinfo {author} {\bibfnamefont {E.}~\bibnamefont {Berti}},\ and\
  \bibinfo {author} {\bibfnamefont {L.}~\bibnamefont {Gualtieri}},\ }\href
  {https://doi.org/10.1103/PhysRevD.88.064048} {\bibfield  {journal} {\bibinfo
  {journal} {Phys. Rev. D}\ }\textbf {\bibinfo {volume} {88}},\ \bibinfo
  {pages} {064048} (\bibinfo {year} {2013}{\natexlab{b}})},\ \Eprint
  {https://arxiv.org/abs/1307.7315} {arXiv:1307.7315 [gr-qc]} \BibitemShut
  {NoStop}%
\bibitem [{\citenamefont {Prasobh}\ and\ \citenamefont
  {Kuriakose}(2014)}]{Prasobh:2014zea}%
  \BibitemOpen
  \bibfield  {author} {\bibinfo {author} {\bibfnamefont {C.~B.}\ \bibnamefont
  {Prasobh}}\ and\ \bibinfo {author} {\bibfnamefont {V.~C.}\ \bibnamefont
  {Kuriakose}},\ }\href {https://doi.org/10.1140/epjc/s10052-014-3136-4}
  {\bibfield  {journal} {\bibinfo  {journal} {Eur. Phys. J. C}\ }\textbf
  {\bibinfo {volume} {74}},\ \bibinfo {pages} {3136} (\bibinfo {year}
  {2014})},\ \Eprint {https://arxiv.org/abs/1405.5334} {arXiv:1405.5334
  [gr-qc]} \BibitemShut {NoStop}%
\bibitem [{\citenamefont {Bhattacharyya}\ and\ \citenamefont
  {Shankaranarayanan}(2017)}]{Bhattacharyya:2017tyc}%
  \BibitemOpen
  \bibfield  {author} {\bibinfo {author} {\bibfnamefont {S.}~\bibnamefont
  {Bhattacharyya}}\ and\ \bibinfo {author} {\bibfnamefont {S.}~\bibnamefont
  {Shankaranarayanan}},\ }\href {https://doi.org/10.1103/PhysRevD.96.064044}
  {\bibfield  {journal} {\bibinfo  {journal} {Phys. Rev. D}\ }\textbf {\bibinfo
  {volume} {96}},\ \bibinfo {pages} {064044} (\bibinfo {year} {2017})},\
  \Eprint {https://arxiv.org/abs/1704.07044} {arXiv:1704.07044 [gr-qc]}
  \BibitemShut {NoStop}%
\bibitem [{\citenamefont {Bhattacharyya}\ and\ \citenamefont
  {Shankaranarayanan}(2019)}]{Bhattacharyya:2018hsj}%
  \BibitemOpen
  \bibfield  {author} {\bibinfo {author} {\bibfnamefont {S.}~\bibnamefont
  {Bhattacharyya}}\ and\ \bibinfo {author} {\bibfnamefont {S.}~\bibnamefont
  {Shankaranarayanan}},\ }\href {https://doi.org/10.1103/PhysRevD.100.024022}
  {\bibfield  {journal} {\bibinfo  {journal} {Phys. Rev. D}\ }\textbf {\bibinfo
  {volume} {100}},\ \bibinfo {pages} {024022} (\bibinfo {year} {2019})},\
  \Eprint {https://arxiv.org/abs/1812.00187} {arXiv:1812.00187 [gr-qc]}
  \BibitemShut {NoStop}%
\bibitem [{\citenamefont {Cruz}\ \emph {et~al.}(2020)\citenamefont {Cruz},
  \citenamefont {Brito},\ and\ \citenamefont {Silva}}]{Cruz:2020emz}%
  \BibitemOpen
  \bibfield  {author} {\bibinfo {author} {\bibfnamefont {M.~B.}\ \bibnamefont
  {Cruz}}, \bibinfo {author} {\bibfnamefont {F.~A.}\ \bibnamefont {Brito}},\
  and\ \bibinfo {author} {\bibfnamefont {C.~A.~S.}\ \bibnamefont {Silva}},\
  }\href {https://doi.org/10.1103/PhysRevD.102.044063} {\bibfield  {journal}
  {\bibinfo  {journal} {Phys. Rev. D}\ }\textbf {\bibinfo {volume} {102}},\
  \bibinfo {pages} {044063} (\bibinfo {year} {2020})},\ \Eprint
  {https://arxiv.org/abs/2005.02208} {arXiv:2005.02208 [gr-qc]} \BibitemShut
  {NoStop}%
\bibitem [{\citenamefont {Chen}\ and\ \citenamefont
  {Park}(2021)}]{Chen:2021pxd}%
  \BibitemOpen
  \bibfield  {author} {\bibinfo {author} {\bibfnamefont {C.-Y.}\ \bibnamefont
  {Chen}}\ and\ \bibinfo {author} {\bibfnamefont {S.}~\bibnamefont {Park}},\
  }\href {https://doi.org/10.1103/PhysRevD.103.064029} {\bibfield  {journal}
  {\bibinfo  {journal} {Phys. Rev. D}\ }\textbf {\bibinfo {volume} {103}},\
  \bibinfo {pages} {064029} (\bibinfo {year} {2021})},\ \Eprint
  {https://arxiv.org/abs/2101.06600} {arXiv:2101.06600 [gr-qc]} \BibitemShut
  {NoStop}%
\bibitem [{\citenamefont {del Corral}\ and\ \citenamefont
  {Olmedo}(2022)}]{del-Corral:2022kbk}%
  \BibitemOpen
  \bibfield  {author} {\bibinfo {author} {\bibfnamefont {D.}~\bibnamefont {del
  Corral}}\ and\ \bibinfo {author} {\bibfnamefont {J.}~\bibnamefont {Olmedo}},\
  }\href {https://doi.org/10.1103/PhysRevD.105.064053} {\bibfield  {journal}
  {\bibinfo  {journal} {Phys. Rev. D}\ }\textbf {\bibinfo {volume} {105}},\
  \bibinfo {pages} {064053} (\bibinfo {year} {2022})},\ \Eprint
  {https://arxiv.org/abs/2201.09584} {arXiv:2201.09584 [gr-qc]} \BibitemShut
  {NoStop}%
\bibitem [{\citenamefont {Cardoso}\ and\ \citenamefont
  {Lemos}(2001)}]{Cardoso:2001bb}%
  \BibitemOpen
  \bibfield  {author} {\bibinfo {author} {\bibfnamefont {V.}~\bibnamefont
  {Cardoso}}\ and\ \bibinfo {author} {\bibfnamefont {J.~P.~S.}\ \bibnamefont
  {Lemos}},\ }\href {https://doi.org/10.1103/PhysRevD.64.084017} {\bibfield
  {journal} {\bibinfo  {journal} {Phys. Rev. D}\ }\textbf {\bibinfo {volume}
  {64}},\ \bibinfo {pages} {084017} (\bibinfo {year} {2001})},\ \Eprint
  {https://arxiv.org/abs/gr-qc/0105103} {arXiv:gr-qc/0105103} \BibitemShut
  {NoStop}%
\bibitem [{\citenamefont {Nunez}\ and\ \citenamefont
  {Starinets}(2003)}]{Nunez:2003eq}%
  \BibitemOpen
  \bibfield  {author} {\bibinfo {author} {\bibfnamefont {A.}~\bibnamefont
  {Nunez}}\ and\ \bibinfo {author} {\bibfnamefont {A.~O.}\ \bibnamefont
  {Starinets}},\ }\href {https://doi.org/10.1103/PhysRevD.67.124013} {\bibfield
   {journal} {\bibinfo  {journal} {Phys. Rev. D}\ }\textbf {\bibinfo {volume}
  {67}},\ \bibinfo {pages} {124013} (\bibinfo {year} {2003})},\ \Eprint
  {https://arxiv.org/abs/hep-th/0302026} {arXiv:hep-th/0302026} \BibitemShut
  {NoStop}%
\bibitem [{\citenamefont {Michalogiorgakis}\ and\ \citenamefont
  {Pufu}(2007)}]{Michalogiorgakis:2006jc}%
  \BibitemOpen
  \bibfield  {author} {\bibinfo {author} {\bibfnamefont {G.}~\bibnamefont
  {Michalogiorgakis}}\ and\ \bibinfo {author} {\bibfnamefont {S.~S.}\
  \bibnamefont {Pufu}},\ }\href {https://doi.org/10.1088/1126-6708/2007/02/023}
  {\bibfield  {journal} {\bibinfo  {journal} {JHEP}\ }\textbf {\bibinfo
  {volume} {02}}\bibfield  {number} {\bibinfo  {number} { (2007)},\ \bibinfo
  {pages} {023}},\ }\Eprint {https://arxiv.org/abs/hep-th/0612065}
  {arXiv:hep-th/0612065} \BibitemShut {NoStop}%
\bibitem [{\citenamefont {Morgan}\ \emph {et~al.}(2009)\citenamefont {Morgan},
  \citenamefont {Cardoso}, \citenamefont {Miranda}, \citenamefont {Molina},\
  and\ \citenamefont {Zanchin}}]{Morgan:2009pn}%
  \BibitemOpen
  \bibfield  {author} {\bibinfo {author} {\bibfnamefont {J.}~\bibnamefont
  {Morgan}}, \bibinfo {author} {\bibfnamefont {V.}~\bibnamefont {Cardoso}},
  \bibinfo {author} {\bibfnamefont {A.~S.}\ \bibnamefont {Miranda}}, \bibinfo
  {author} {\bibfnamefont {C.}~\bibnamefont {Molina}},\ and\ \bibinfo {author}
  {\bibfnamefont {V.~T.}\ \bibnamefont {Zanchin}},\ }\href
  {https://doi.org/10.1088/1126-6708/2009/09/117} {\bibfield  {journal}
  {\bibinfo  {journal} {JHEP}\ }\textbf {\bibinfo {volume} {09}}\bibfield
  {number} {\bibinfo  {number} { (2009)},\ \bibinfo {pages} {117}},\ }\Eprint
  {https://arxiv.org/abs/0907.5011} {arXiv:0907.5011 [hep-th]} \BibitemShut
  {NoStop}%
\bibitem [{\citenamefont {Gambini}\ \emph {et~al.}(2022)\citenamefont
  {Gambini}, \citenamefont {Ben\'\i{}tez},\ and\ \citenamefont
  {Pullin}}]{Gambini:2021uzf}%
  \BibitemOpen
  \bibfield  {author} {\bibinfo {author} {\bibfnamefont {R.}~\bibnamefont
  {Gambini}}, \bibinfo {author} {\bibfnamefont {F.}~\bibnamefont
  {Ben\'\i{}tez}},\ and\ \bibinfo {author} {\bibfnamefont {J.}~\bibnamefont
  {Pullin}},\ }\href {https://doi.org/10.3390/universe8100526} {\bibfield
  {journal} {\bibinfo  {journal} {Universe}\ }\textbf {\bibinfo {volume} {8}},\
  \bibinfo {pages} {526} (\bibinfo {year} {2022})},\ \Eprint
  {https://arxiv.org/abs/2102.09501} {arXiv:2102.09501 [gr-qc]} \BibitemShut
  {NoStop}%
\bibitem [{\citenamefont {Alonso-Bardaji}\ and\ \citenamefont
  {Brizuela}(2021)}]{Alonso-Bardaji:2021tvy}%
  \BibitemOpen
  \bibfield  {author} {\bibinfo {author} {\bibfnamefont {A.}~\bibnamefont
  {Alonso-Bardaji}}\ and\ \bibinfo {author} {\bibfnamefont {D.}~\bibnamefont
  {Brizuela}},\ }\href {https://doi.org/10.1103/PhysRevD.104.084064} {\bibfield
   {journal} {\bibinfo  {journal} {Phys. Rev. D}\ }\textbf {\bibinfo {volume}
  {104}},\ \bibinfo {pages} {084064} (\bibinfo {year} {2021})},\ \Eprint
  {https://arxiv.org/abs/2106.07595} {arXiv:2106.07595 [gr-qc]} \BibitemShut
  {NoStop}%
\bibitem [{\citenamefont {Bojowald}\ \emph {et~al.}(2018)\citenamefont
  {Bojowald}, \citenamefont {Brahma},\ and\ \citenamefont
  {Yeom}}]{Bojowald:2018xxu}%
  \BibitemOpen
  \bibfield  {author} {\bibinfo {author} {\bibfnamefont {M.}~\bibnamefont
  {Bojowald}}, \bibinfo {author} {\bibfnamefont {S.}~\bibnamefont {Brahma}},\
  and\ \bibinfo {author} {\bibfnamefont {D.-h.}\ \bibnamefont {Yeom}},\ }\href
  {https://doi.org/10.1103/PhysRevD.98.046015} {\bibfield  {journal} {\bibinfo
  {journal} {Phys. Rev. D}\ }\textbf {\bibinfo {volume} {98}},\ \bibinfo
  {pages} {046015} (\bibinfo {year} {2018})},\ \Eprint
  {https://arxiv.org/abs/1803.01119} {arXiv:1803.01119 [gr-qc]} \BibitemShut
  {NoStop}%
\bibitem [{\citenamefont {Bojowald}(2020)}]{Bojowald:2020unm}%
  \BibitemOpen
  \bibfield  {author} {\bibinfo {author} {\bibfnamefont {M.}~\bibnamefont
  {Bojowald}},\ }\href {https://doi.org/10.1103/PhysRevD.102.046006} {\bibfield
   {journal} {\bibinfo  {journal} {Phys. Rev. D}\ }\textbf {\bibinfo {volume}
  {102}},\ \bibinfo {pages} {046006} (\bibinfo {year} {2020})},\ \Eprint
  {https://arxiv.org/abs/2007.16066} {arXiv:2007.16066 [gr-qc]} \BibitemShut
  {NoStop}%
\bibitem [{\citenamefont {Borges}\ \emph {et~al.}(2024)\citenamefont {Borges},
  \citenamefont {Baranov}, \citenamefont {Sobrinho},\ and\ \citenamefont
  {Carneiro}}]{Borges:2023fub}%
  \BibitemOpen
  \bibfield  {author} {\bibinfo {author} {\bibfnamefont {H.~A.}\ \bibnamefont
  {Borges}}, \bibinfo {author} {\bibfnamefont {I.~P.~R.}\ \bibnamefont
  {Baranov}}, \bibinfo {author} {\bibfnamefont {F.~C.}\ \bibnamefont
  {Sobrinho}},\ and\ \bibinfo {author} {\bibfnamefont {S.}~\bibnamefont
  {Carneiro}},\ }\href {https://doi.org/10.1088/1361-6382/ad210c} {\bibfield
  {journal} {\bibinfo  {journal} {Class. Quant. Grav.}\ }\textbf {\bibinfo
  {volume} {41}},\ \bibinfo {pages} {05LT01} (\bibinfo {year} {2024})},\
  \Eprint {https://arxiv.org/abs/2310.01560} {arXiv:2310.01560 [gr-qc]}
  \BibitemShut {NoStop}%
\bibitem [{\citenamefont {Chen}\ and\ \citenamefont
  {Chen}(2019)}]{Chen:2019iuo}%
  \BibitemOpen
  \bibfield  {author} {\bibinfo {author} {\bibfnamefont {C.-Y.}\ \bibnamefont
  {Chen}}\ and\ \bibinfo {author} {\bibfnamefont {P.}~\bibnamefont {Chen}},\
  }\href {https://doi.org/10.1103/PhysRevD.99.104003} {\bibfield  {journal}
  {\bibinfo  {journal} {Phys. Rev. D}\ }\textbf {\bibinfo {volume} {99}},\
  \bibinfo {pages} {104003} (\bibinfo {year} {2019})},\ \Eprint
  {https://arxiv.org/abs/1902.01678} {arXiv:1902.01678 [gr-qc]} \BibitemShut
  {NoStop}%
\bibitem [{\citenamefont {Yang}\ \emph {et~al.}(2023)\citenamefont {Yang},
  \citenamefont {Guo}, \citenamefont {Tan},\ and\ \citenamefont
  {Liu}}]{Yang:2023gas}%
  \BibitemOpen
  \bibfield  {author} {\bibinfo {author} {\bibfnamefont {S.}~\bibnamefont
  {Yang}}, \bibinfo {author} {\bibfnamefont {W.-D.}\ \bibnamefont {Guo}},
  \bibinfo {author} {\bibfnamefont {Q.}~\bibnamefont {Tan}},\ and\ \bibinfo
  {author} {\bibfnamefont {Y.-X.}\ \bibnamefont {Liu}},\ }\href
  {https://doi.org/10.1103/PhysRevD.108.024055} {\bibfield  {journal} {\bibinfo
   {journal} {Phys. Rev. D}\ }\textbf {\bibinfo {volume} {108}},\ \bibinfo
  {pages} {024055} (\bibinfo {year} {2023})},\ \Eprint
  {https://arxiv.org/abs/2304.06895} {arXiv:2304.06895 [gr-qc]} \BibitemShut
  {NoStop}%
\bibitem [{\citenamefont {Arbey}\ \emph {et~al.}(2021)\citenamefont {Arbey},
  \citenamefont {Auffinger}, \citenamefont {Geiller}, \citenamefont {Livine},\
  and\ \citenamefont {Sartini}}]{Arbey:2021jif}%
  \BibitemOpen
  \bibfield  {author} {\bibinfo {author} {\bibfnamefont {A.}~\bibnamefont
  {Arbey}}, \bibinfo {author} {\bibfnamefont {J.}~\bibnamefont {Auffinger}},
  \bibinfo {author} {\bibfnamefont {M.}~\bibnamefont {Geiller}}, \bibinfo
  {author} {\bibfnamefont {E.~R.}\ \bibnamefont {Livine}},\ and\ \bibinfo
  {author} {\bibfnamefont {F.}~\bibnamefont {Sartini}},\ }\href
  {https://doi.org/10.1103/PhysRevD.103.104010} {\bibfield  {journal} {\bibinfo
   {journal} {Phys. Rev. D}\ }\textbf {\bibinfo {volume} {103}},\ \bibinfo
  {pages} {104010} (\bibinfo {year} {2021})},\ \Eprint
  {https://arxiv.org/abs/2101.02951} {arXiv:2101.02951 [gr-qc]} \BibitemShut
  {NoStop}%
\bibitem [{\citenamefont {Hossenfelder}\ \emph {et~al.}(2012)\citenamefont
  {Hossenfelder}, \citenamefont {Modesto},\ and\ \citenamefont
  {Premont-Schwarz}}]{Hossenfelder:2012tc}%
  \BibitemOpen
  \bibfield  {author} {\bibinfo {author} {\bibfnamefont {S.}~\bibnamefont
  {Hossenfelder}}, \bibinfo {author} {\bibfnamefont {L.}~\bibnamefont
  {Modesto}},\ and\ \bibinfo {author} {\bibfnamefont {I.}~\bibnamefont
  {Premont-Schwarz}},\ }\href@noop {} {\bibfield  {journal} {\bibinfo
  {journal} {arXiv:1202.0412}\ } (\bibinfo {year} {2012})},\ \Eprint
  {https://arxiv.org/abs/1202.0412} {arXiv:1202.0412 [gr-qc]} \BibitemShut
  {NoStop}%
\bibitem [{\citenamefont {Moulin}\ \emph {et~al.}(2019)\citenamefont {Moulin},
  \citenamefont {Barrau},\ and\ \citenamefont {Martineau}}]{Moulin:2019ekf}%
  \BibitemOpen
  \bibfield  {author} {\bibinfo {author} {\bibfnamefont {F.}~\bibnamefont
  {Moulin}}, \bibinfo {author} {\bibfnamefont {A.}~\bibnamefont {Barrau}},\
  and\ \bibinfo {author} {\bibfnamefont {K.}~\bibnamefont {Martineau}},\ }\href
  {https://doi.org/10.3390/universe5090202} {\bibfield  {journal} {\bibinfo
  {journal} {Universe}\ }\textbf {\bibinfo {volume} {5}},\ \bibinfo {pages}
  {202} (\bibinfo {year} {2019})},\ \Eprint {https://arxiv.org/abs/1908.06311}
  {arXiv:1908.06311 [gr-qc]} \BibitemShut {NoStop}%
\bibitem [{\citenamefont {{Schutz}}\ and\ \citenamefont
  {{Will}}(1985)}]{1985ApJ...291L..33S}%
  \BibitemOpen
  \bibfield  {author} {\bibinfo {author} {\bibfnamefont {B.~F.}\ \bibnamefont
  {{Schutz}}}\ and\ \bibinfo {author} {\bibfnamefont {C.~M.}\ \bibnamefont
  {{Will}}},\ }\href {https://doi.org/10.1086/184453} {\bibfield  {journal}
  {\bibinfo  {journal} {Astrophys. J.}\ }\textbf {\bibinfo {volume} {35}},\
  \bibinfo {pages} {3621} (\bibinfo {year} {1985})}\BibitemShut {NoStop}%
\bibitem [{\citenamefont {Iyer}\ and\ \citenamefont
  {Will}(1987)}]{PhysRevD.35.3621}%
  \BibitemOpen
  \bibfield  {author} {\bibinfo {author} {\bibfnamefont {S.}~\bibnamefont
  {Iyer}}\ and\ \bibinfo {author} {\bibfnamefont {C.~M.}\ \bibnamefont
  {Will}},\ }\href {https://doi.org/10.1103/PhysRevD.35.3621} {\bibfield
  {journal} {\bibinfo  {journal} {Phys. Rev. D}\ }\textbf {\bibinfo {volume}
  {35}},\ \bibinfo {pages} {3621} (\bibinfo {year} {1987})}\BibitemShut
  {NoStop}%
\bibitem [{\citenamefont {Matyjasek}\ and\ \citenamefont
  {Opala}(2017)}]{Matyjasek:2017psv}%
  \BibitemOpen
  \bibfield  {author} {\bibinfo {author} {\bibfnamefont {J.}~\bibnamefont
  {Matyjasek}}\ and\ \bibinfo {author} {\bibfnamefont {M.}~\bibnamefont
  {Opala}},\ }\href {https://doi.org/10.1103/PhysRevD.96.024011} {\bibfield
  {journal} {\bibinfo  {journal} {Phys. Rev. D}\ }\textbf {\bibinfo {volume}
  {96}},\ \bibinfo {pages} {024011} (\bibinfo {year} {2017})},\ \Eprint
  {https://arxiv.org/abs/1704.00361} {arXiv:1704.00361 [gr-qc]} \BibitemShut
  {NoStop}%
\bibitem [{\citenamefont {Konoplya}\ \emph {et~al.}(2019)\citenamefont
  {Konoplya}, \citenamefont {Zhidenko},\ and\ \citenamefont
  {Zinhailo}}]{Konoplya:2019hlu}%
  \BibitemOpen
  \bibfield  {author} {\bibinfo {author} {\bibfnamefont {R.~A.}\ \bibnamefont
  {Konoplya}}, \bibinfo {author} {\bibfnamefont {A.}~\bibnamefont {Zhidenko}},\
  and\ \bibinfo {author} {\bibfnamefont {A.~F.}\ \bibnamefont {Zinhailo}},\
  }\href {https://doi.org/10.1088/1361-6382/ab2e25} {\bibfield  {journal}
  {\bibinfo  {journal} {Class. Quant. Grav.}\ }\textbf {\bibinfo {volume}
  {36}},\ \bibinfo {pages} {155002} (\bibinfo {year} {2019})},\ \Eprint
  {https://arxiv.org/abs/1904.10333} {arXiv:1904.10333 [gr-qc]} \BibitemShut
  {NoStop}%
\bibitem [{\citenamefont {Mamani}\ \emph {et~al.}(2022)\citenamefont {Mamani},
  \citenamefont {Masa}, \citenamefont {Sanches},\ and\ \citenamefont
  {Zanchin}}]{Mamani:2022akq}%
  \BibitemOpen
  \bibfield  {author} {\bibinfo {author} {\bibfnamefont {L.~A.~H.}\
  \bibnamefont {Mamani}}, \bibinfo {author} {\bibfnamefont {A.~D.~D.}\
  \bibnamefont {Masa}}, \bibinfo {author} {\bibfnamefont {L.~T.}\ \bibnamefont
  {Sanches}},\ and\ \bibinfo {author} {\bibfnamefont {V.~T.}\ \bibnamefont
  {Zanchin}},\ }\href {https://doi.org/10.1140/epjc/s10052-022-10865-1}
  {\bibfield  {journal} {\bibinfo  {journal} {Eur. Phys. J. C}\ }\textbf
  {\bibinfo {volume} {82}},\ \bibinfo {pages} {897} (\bibinfo {year} {2022})},\
  \Eprint {https://arxiv.org/abs/2206.03512} {arXiv:2206.03512 [gr-qc]}
  \BibitemShut {NoStop}%
\bibitem [{\citenamefont {Jansen}(2017)}]{Jansen:2017oag}%
  \BibitemOpen
  \bibfield  {author} {\bibinfo {author} {\bibfnamefont {A.}~\bibnamefont
  {Jansen}},\ }\href {https://doi.org/10.1140/epjp/i2017-11825-9} {\bibfield
  {journal} {\bibinfo  {journal} {Eur. Phys. J. Plus}\ }\textbf {\bibinfo
  {volume} {132}},\ \bibinfo {pages} {546} (\bibinfo {year} {2017})},\ \Eprint
  {https://arxiv.org/abs/1709.09178} {arXiv:1709.09178 [gr-qc]} \BibitemShut
  {NoStop}%
\bibitem [{\citenamefont {Leaver}(1985)}]{Leaver:1985ax}%
  \BibitemOpen
  \bibfield  {author} {\bibinfo {author} {\bibfnamefont {E.~W.}\ \bibnamefont
  {Leaver}},\ }\href {https://doi.org/10.1098/rspa.1985.0119} {\bibfield
  {journal} {\bibinfo  {journal} {Proc. Roy. Soc. Lond. A}\ }\textbf {\bibinfo
  {volume} {402}},\ \bibinfo {pages} {285} (\bibinfo {year}
  {1985})}\BibitemShut {NoStop}%
\bibitem [{\citenamefont {Leaver}(1990)}]{PhysRevD.41.2986}%
  \BibitemOpen
  \bibfield  {author} {\bibinfo {author} {\bibfnamefont {E.~W.}\ \bibnamefont
  {Leaver}},\ }\href {https://doi.org/10.1103/PhysRevD.41.2986} {\bibfield
  {journal} {\bibinfo  {journal} {Phys. Rev. D}\ }\textbf {\bibinfo {volume}
  {41}},\ \bibinfo {pages} {2986} (\bibinfo {year} {1990})}\BibitemShut
  {NoStop}%
\bibitem [{\citenamefont {Nollert}(1993)}]{PhysRevD.47.5253}%
  \BibitemOpen
  \bibfield  {author} {\bibinfo {author} {\bibfnamefont {H.-P.}\ \bibnamefont
  {Nollert}},\ }\href {https://doi.org/10.1103/PhysRevD.47.5253} {\bibfield
  {journal} {\bibinfo  {journal} {Phys. Rev. D}\ }\textbf {\bibinfo {volume}
  {47}},\ \bibinfo {pages} {5253} (\bibinfo {year} {1993})}\BibitemShut
  {NoStop}%
\bibitem [{\citenamefont {Lentz}(1976)}]{Lentz:76}%
  \BibitemOpen
  \bibfield  {author} {\bibinfo {author} {\bibfnamefont {W.~J.}\ \bibnamefont
  {Lentz}},\ }\href {https://doi.org/10.1364/AO.15.000668} {\bibfield
  {journal} {\bibinfo  {journal} {Appl. Opt.}\ }\textbf {\bibinfo {volume}
  {15}},\ \bibinfo {pages} {668} (\bibinfo {year} {1976})}\BibitemShut
  {NoStop}%
\bibitem [{\citenamefont {Jing}\ and\ \citenamefont {Pan}(2008)}]{Jing:2008an}%
  \BibitemOpen
  \bibfield  {author} {\bibinfo {author} {\bibfnamefont {J.}~\bibnamefont
  {Jing}}\ and\ \bibinfo {author} {\bibfnamefont {Q.}~\bibnamefont {Pan}},\
  }\href {https://doi.org/10.1016/j.physletb.2007.11.039} {\bibfield  {journal}
  {\bibinfo  {journal} {Phys. Lett. B}\ }\textbf {\bibinfo {volume} {660}},\
  \bibinfo {pages} {13} (\bibinfo {year} {2008})},\ \Eprint
  {https://arxiv.org/abs/0802.0043} {arXiv:0802.0043 [gr-qc]} \BibitemShut
  {NoStop}%
\bibitem [{\citenamefont {Berti}\ and\ \citenamefont
  {Kokkotas}(2003)}]{Berti:2003zu}%
  \BibitemOpen
  \bibfield  {author} {\bibinfo {author} {\bibfnamefont {E.}~\bibnamefont
  {Berti}}\ and\ \bibinfo {author} {\bibfnamefont {K.~D.}\ \bibnamefont
  {Kokkotas}},\ }\href {https://doi.org/10.1103/PhysRevD.68.044027} {\bibfield
  {journal} {\bibinfo  {journal} {Phys. Rev. D}\ }\textbf {\bibinfo {volume}
  {68}},\ \bibinfo {pages} {044027} (\bibinfo {year} {2003})},\ \Eprint
  {https://arxiv.org/abs/hep-th/0303029} {arXiv:hep-th/0303029} \BibitemShut
  {NoStop}%
\bibitem [{\citenamefont {Konoplya}\ \emph {et~al.}(2022)\citenamefont
  {Konoplya}, \citenamefont {Zinhailo}, \citenamefont {Kunz}, \citenamefont
  {Stuchlik},\ and\ \citenamefont {Zhidenko}}]{Konoplya:2022hll}%
  \BibitemOpen
  \bibfield  {author} {\bibinfo {author} {\bibfnamefont {R.~A.}\ \bibnamefont
  {Konoplya}}, \bibinfo {author} {\bibfnamefont {A.~F.}\ \bibnamefont
  {Zinhailo}}, \bibinfo {author} {\bibfnamefont {J.}~\bibnamefont {Kunz}},
  \bibinfo {author} {\bibfnamefont {Z.}~\bibnamefont {Stuchlik}},\ and\
  \bibinfo {author} {\bibfnamefont {A.}~\bibnamefont {Zhidenko}},\ }\href
  {https://doi.org/10.1088/1475-7516/2022/10/091} {\bibfield  {journal}
  {\bibinfo  {journal} {JCAP}\ }\textbf {\bibinfo {volume} {10}}\bibfield
  {number} {\bibinfo  {number} { (091)}},\ }\Eprint
  {https://arxiv.org/abs/2206.14714} {arXiv:2206.14714 [gr-qc]} \BibitemShut
  {NoStop}%
\bibitem [{\citenamefont {Zhang}\ \emph {et~al.}(2024)\citenamefont {Zhang},
  \citenamefont {Gong}, \citenamefont {Fu}, \citenamefont {Wu},\ and\
  \citenamefont {Pan}}]{Zhang:2024nny}%
  \BibitemOpen
  \bibfield  {author} {\bibinfo {author} {\bibfnamefont {D.}~\bibnamefont
  {Zhang}}, \bibinfo {author} {\bibfnamefont {H.}~\bibnamefont {Gong}},
  \bibinfo {author} {\bibfnamefont {G.}~\bibnamefont {Fu}}, \bibinfo {author}
  {\bibfnamefont {J.-P.}\ \bibnamefont {Wu}},\ and\ \bibinfo {author}
  {\bibfnamefont {Q.}~\bibnamefont {Pan}},\ }\href
  {https://doi.org/10.1140/epjc/s10052-024-12928-x} {\bibfield  {journal}
  {\bibinfo  {journal} {Eur. Phys. J. C}\ }\textbf {\bibinfo {volume} {84}},\
  \bibinfo {pages} {564} (\bibinfo {year} {2024})},\ \Eprint
  {https://arxiv.org/abs/2402.15085} {arXiv:2402.15085 [gr-qc]} \BibitemShut
  {NoStop}%
\bibitem [{\citenamefont {Konoplya}\ and\ \citenamefont
  {Zhidenko}(2022)}]{Konoplya:2022pbc}%
  \BibitemOpen
  \bibfield  {author} {\bibinfo {author} {\bibfnamefont {R.~A.}\ \bibnamefont
  {Konoplya}}\ and\ \bibinfo {author} {\bibfnamefont {A.}~\bibnamefont
  {Zhidenko}},\ }\href@noop {} {\bibfield  {journal} {\bibinfo  {journal}
  {arXiv:2209.00679}\ } (\bibinfo {year} {2022})},\ \Eprint
  {https://arxiv.org/abs/2209.00679} {arXiv:2209.00679 [gr-qc]} \BibitemShut
  {NoStop}%
\bibitem [{\citenamefont {Moulin}\ and\ \citenamefont
  {Barrau}(2020)}]{Moulin:2019bfh}%
  \BibitemOpen
  \bibfield  {author} {\bibinfo {author} {\bibfnamefont {F.}~\bibnamefont
  {Moulin}}\ and\ \bibinfo {author} {\bibfnamefont {A.}~\bibnamefont
  {Barrau}},\ }\href {https://doi.org/10.1007/s10714-020-02737-4} {\bibfield
  {journal} {\bibinfo  {journal} {Gen. Rel. Grav.}\ }\textbf {\bibinfo {volume}
  {52}},\ \bibinfo {pages} {82} (\bibinfo {year} {2020})},\ \Eprint
  {https://arxiv.org/abs/1906.05633} {arXiv:1906.05633 [gr-qc]} \BibitemShut
  {NoStop}%
\bibitem [{\citenamefont {Bouhmadi-L\'opez}\ \emph {et~al.}(2020)\citenamefont
  {Bouhmadi-L\'opez}, \citenamefont {Brahma}, \citenamefont {Chen},
  \citenamefont {Chen},\ and\ \citenamefont {Yeom}}]{Bouhmadi-Lopez:2020oia}%
  \BibitemOpen
  \bibfield  {author} {\bibinfo {author} {\bibfnamefont {M.}~\bibnamefont
  {Bouhmadi-L\'opez}}, \bibinfo {author} {\bibfnamefont {S.}~\bibnamefont
  {Brahma}}, \bibinfo {author} {\bibfnamefont {C.-Y.~u.}\ \bibnamefont {Chen}},
  \bibinfo {author} {\bibfnamefont {P.}~\bibnamefont {Chen}},\ and\ \bibinfo
  {author} {\bibfnamefont {D.-h.}\ \bibnamefont {Yeom}},\ }\href
  {https://doi.org/10.1088/1475-7516/2020/07/066} {\bibfield  {journal}
  {\bibinfo  {journal} {JCAP}\ }\textbf {\bibinfo {volume} {07}}\bibfield
  {number} {\bibinfo  {number} { (066)}},\ }\Eprint
  {https://arxiv.org/abs/2004.13061} {arXiv:2004.13061 [gr-qc]} \BibitemShut
  {NoStop}%
\end{thebibliography}%
\end{document}